\documentclass[sigconf,letterpaper,9pt]{acmart}

  {\list{}{\leftmargin=0.3in\rightmargin=0.3in}\item[]}%
  {\endlist}

\usepackage{xcolor}
\usepackage[caption=false,font=normalsize,labelfont=bf,labelsep=period]{subfig}
\usepackage[justification=centering]{caption}
\usepackage{soul}
\usepackage[noend]{algpseudocode}
\usepackage{multirow}
\usepackage{algorithm}
\usepackage{xspace}
\usepackage{tikz}
\usetikzlibrary{arrows.meta, automata, positioning, shapes.geometric, backgrounds, fit, calc}
\usepackage{paralist}
\usepackage{colortbl}
\usepackage{amsmath}
\usepackage{tabularx}
\usepackage{array}
\usepackage{makecell, cellspace}
\usepackage{comment}
\usepackage{url}
\usepackage{float}
\usepackage{graphicx}
\usepackage{mdframed}
\usepackage{placeins}
\usepackage{booktabs}
\usepackage{lipsum}
\usepackage{multicol}
\usepackage{bmpsize}
\usepackage{textcomp}
\usepackage{times}

\PassOptionsToPackage{x11names}{xcolor}
\usetikzlibrary{arrows.meta, automata, positioning, shapes.geometric, backgrounds, fit, calc}
\newcommand{\system}{{\ensuremath{\sf{TRAIN}}}\xspace}
\newcommand{\trapsrtc}{{\ensuremath{\sf{TRAIN_{A}}}}\xspace}
\newcommand{\trapsnortc}{{\ensuremath{\sf{TRAIN_{B}}}}\xspace}

\newcommand{\Attreq}{\textit{Att$_{request}$}}
\newcommand{\Attrep}{\textit{Att$_{report}$}}

\newcommand{\Authrep}{\textit{Auth$_{report}$}}

\newcommand{\snd}{\textit{ID$_{Snd}$}}

\newcommand{\hash}{\textit{Hash$_{New}$}}
\newcommand{\curhash}{\textit{Hash$_{Cur}$}}
\newcommand{\hashind}{\textit {HashInd$_{New}$}}
\newcommand{\curhashind}{\textit {HashInd$_{Cur}$}}
\newcommand{\attesttime}{\textit{t$_{attest}$}}

\newcommand{\timeout}{\textit{t$_{timeout}$}}
\newcommand{\devid}{\textit{ID$_{Dev}$}}
\newcommand{\parent}{\textit{ID$_{Par}$}}
\newcommand{\maxdelay}{\textit{t$_{maxDelay}$}}

\newcommand{\height}{\textit{Height$_{Cur}$}}
\newcommand{\netheight}{\textit{Height$_{Net}$}}
\newcommand{\lmt}{\textit{LMT$_{Dev}$}}

\newcommand{\ra}{{\ensuremath{\sf{\mathcal RA}}}\xspace}
\newcommand{\sa}{{\ensuremath{\sf{\mathcal NA}}}\xspace}
\newcommand{\sadv}{{\ensuremath{\sf{\mathcal Adv}}}\xspace}
\newcommand{\chal}{{\ensuremath{\sf{\mathcal Chal}}}\xspace}
\newcommand{\prv}{{\ensuremath{\sf{\mathcal Prv}}}\xspace}
\newcommand{\vrf}{{\ensuremath{\sf{\mathcal Vrf}}}\xspace}

\newcommand{\toctou}{{\ensuremath{\sf{ TOCTOU}}}\xspace}
\newcommand{\toctoura}{{\ensuremath{\sf{ TOCTOU_{\ra}}}}\xspace}
\newcommand{\toctousa}{{\ensuremath{\sf{ TOCTOU_{\sa}}}}\xspace}
\newcommand{\key}{\ensuremath{\mathcal K}$_{Dev}$\xspace}
\newcommand{\rot}{{{\it RoT}}\xspace}
\newcommand{\casu}{{{\rm CASU}}\xspace}
\newcommand{\rata}{{{\rm RATA}}\xspace}
\newcommand{\garota}{{{\rm GAROTA}}\xspace}
\long\def\comment#1{}
\newcommand{\trapscasu}{{\ensuremath{\sf{TRAIN_\casu}}}\xspace}
\newcommand{\trapsrata}{{\ensuremath{\sf{TRAIN_\rata}}}\xspace}
\newcommand{\pc}{\ensuremath {PC}\xspace}
\newcommand{\er}{\ensuremath {ER}\xspace}
\newcommand{\ep}{\ensuremath {EP}\xspace}
\newcommand{\tcr}{\ensuremath {TCR}\xspace}
\newcommand{\ivtr}{\ensuremath {IVTR}\xspace}
\newcommand{\wen}{\ensuremath{W_{en}}\xspace}
\newcommand{\daddr}{\ensuremath {D_{addr}}\xspace}
\newcommand{\irq}{\ensuremath {irq}\xspace}
\newcommand{\irqcfg}{\ensuremath {IRQ_{cfg}}\xspace}

\newcommand{\dmaen}{\ensuremath {DMA_{en}}\xspace}
\newcommand{\dmaaddr}{\ensuremath {DMA_{addr}}\xspace}
\newcommand{\gie}{\ensuremath {gie}\xspace}
\newcommand{\isrt}{\ensuremath {ISR}_T\xspace}
\newcommand{\isrtmin}{\ensuremath {ISR_{T_{min}}}\xspace}
\newcommand{\isrtmax}{\ensuremath {ISR_{T_{max}}}\xspace}
\newcommand{\isru}{\ensuremath {ISR}_U\xspace}
\newcommand{\isrumin}{\ensuremath {ISR_{U_{min}}}\xspace}
\newcommand{\isrumax}{\ensuremath {ISR_{U_{max}}}\xspace}
\newcommand{\casumem}{\ensuremath {M_{\system}}\xspace}
\newcommand{\reset}{\ensuremath {reset}\xspace}

\usepackage{pifont}
\newcommand{\greencheck}{\textcolor{green}{\ding{51}}}
\newcommand{\greencheckt}{\textcolor{green}{\ding{52}}}
\newcommand{\redcross}{\textcolor{red}{\ding{56}}}
\makeatletter
\newcommand{\thickhline}{%
    \noalign {\ifnum 0=`}\fi \hrule height 1pt
    \futurelet \reserved@a \@xhline
}
\newcolumntype{"}{@{\hskip\tabcolsep\vrule width 1pt\hskip\tabcolsep}}
\makeatother

\newcommand{\systemtext}{{\underline{T}OCTOU \underline{R}esilient \underline{A}ttestation for \underline{I}oT \underline{N}etworks}}
\newcommand{\systemtitle}{{TOCTOU Resilient 
Attestation for IoT Networks \\ (Full Version)}}
\newcommand*{\Scale}[2][4]{\scalebox{#1}{$#2$}}%

\title{\systemtitle}

\author{Pavel Frolikov}
\email{pavel@uci.edu}
\affiliation{%
  \institution{UC Irvine}  
  \state{CA}
  \country{USA}
}

\author{Youngil Kim}
\email{youngik2@uci.edu}
\affiliation{%
  \institution{UC Irvine}  
  \state{CA}
  \country{USA}
}

\author{Renascence Tarafder Prapty}
\email{rprapty@uci.edu}
\affiliation{%
  \institution{UC Irvine}  
  \state{CA}
  \country{USA}
}

\author{Gene Tsudik}
\email{gene.tsudik@uci.edu}
\affiliation{%
  \institution{UC Irvine}  
  \state{CA}
  \country{USA}
}

\setcopyright{none}
\renewcommand\footnotetextcopyrightpermission[1]{} 

\begin{document}
\begin{abstract}
Internet-of-Things (IoT) devices are increasingly common in both consumer and industrial settings, often 
performing safety-critical functions. Although securing these devices is vital, manufacturers typically 
neglect security issues or address them as an afterthought. This is of particular importance
in IoT networks, e.g., in the industrial automation settings.

To this end, network attestation -- verifying the software state of all devices in a network -- is a promising mitigation approach. However, current network attestation schemes have certain shortcomings:
(1) lengthy TOCTOU (Time-Of-Check-Time-Of-Use) vulnerability windows,
(2) high latency and resource overhead, and
(3) susceptibility to interference from compromised devices.
To address these limitations, we construct TRAIN (TOCTOU-Resilient Attestation for IoT Networks), 
an efficient technique that minimizes TOCTOU windows, ensures constant-time per-device attestation, 
and maintains resilience even with multiple compromised devices. We demonstrate TRAIN's viability 
and evaluate its performance via a fully functional and publicly available prototype.

\end{abstract}
\maketitle
\thispagestyle{plain}
\pagestyle{plain}
\vspace{-0.4cm}
\section{Introduction \label{intro}}
Rapid expansion and popularity of the Internet of Things (IoT) devices 
and Cyber-Physical Systems (CPS) have resulted in the deployment of vast numbers of 
Internet-connected and inter-connected devices. Such networks, 
composed of numerous devices, collaboratively execute sensing and/or actuation 
tasks in diverse settings, such as smart factories, warehouses, agriculture,
and environmental monitoring. However, 
the resource-constrained nature of IoT devices makes them vulnerable to 
remote attacks. This poses significant risks: malicious actors 
can compromise data integrity or even jeopardize safety within critical 
control loops. Given the safety-critical functions they perform 
and the sensitive data they collect, protecting IoT devices 
against such attacks is essential.
Remote attestation (\ra), a well-established security service, detects
malware on remote devices by verifying the integrity of their 
software state \cite{nunes2019towards, rata, li2011viper}. However, applying 
single-device \ra techniques to large IoT networks incurs high overhead.
Many techniques, including 
\cite{asokan2015seda,ambrosin2016sana,carpent2017lightweight, kohnhauser2017scapi,petzi2022scraps}, 
made progress towards efficient network  (aka swarm) attestation (\sa).
Nonetheless, they have substantial limitations, which form the motivation for this work.

\noindent{\bf Time-of-Check to Time-of-Use ({\toctou}):}
Prior techniques do not guarantee simultaneous (synchronized) attestation 
across all networked devices. Network structure, potential mobility, intermittent
connectivity, and congestion can lead to staggered reception of \ra requests,  
thus widening the time window for discrepancies in \ra timing.  
Also, even if networked devices are all of the same type, varying memory sizes 
and application task scheduling can result in different execution times of \ra.
These factors lead to a potentially long \toctou window, 
where the state of network devices is captured over an 
interval of time rather than at the same time. This increases the risk of 
undetected transient malware presence.  
\begin{figure}
    \includegraphics [width=\columnwidth]
    {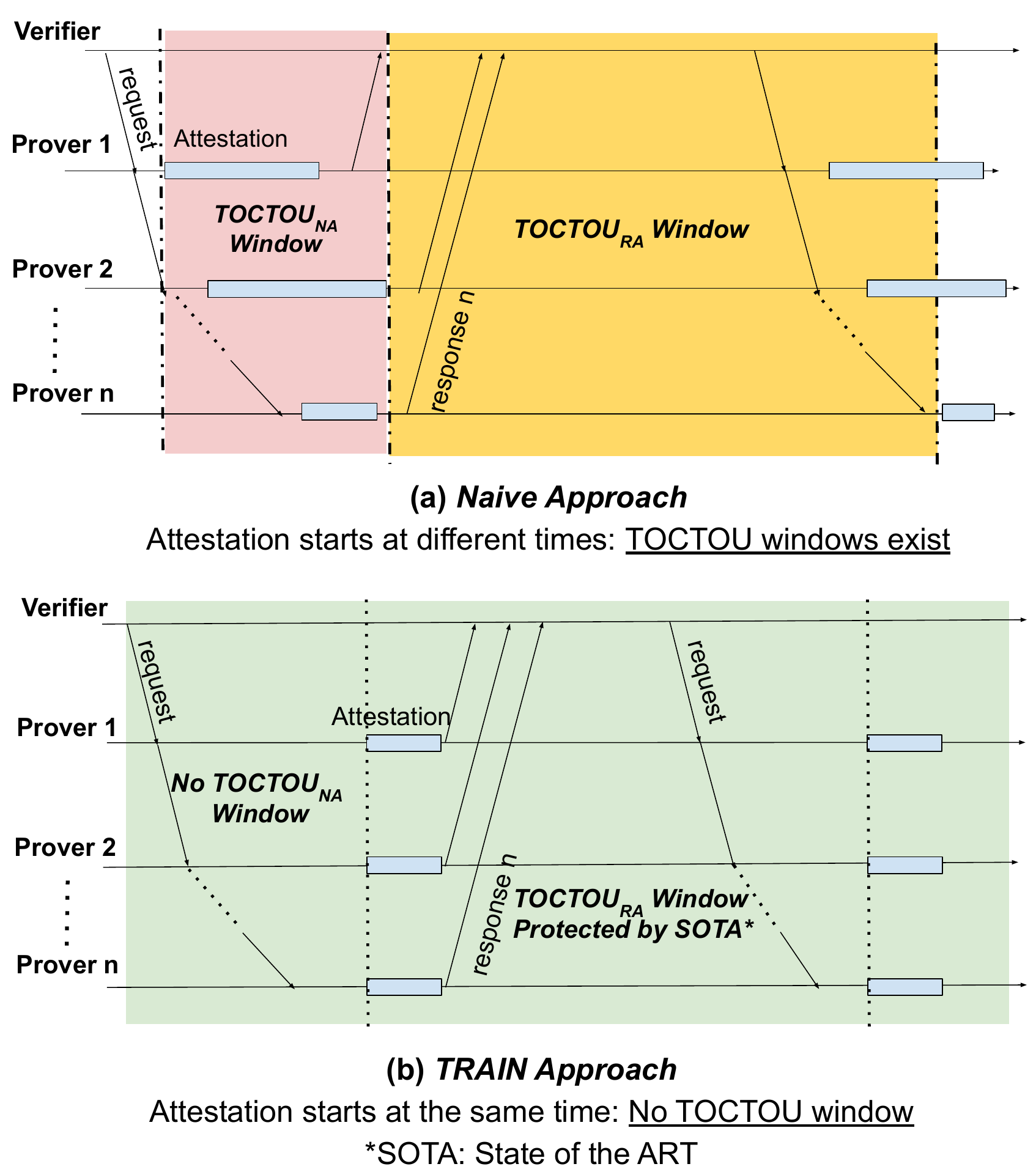}
   \captionsetup{justification=centering}
   \vspace{-1.8em}
   \caption{TOCTOU Window Minimized by \system}
    \label{fig: toctou window}
    \vspace{-1.5em}
\end{figure}
The \toctou problem arises in two cases: 

\noindent $\boldsymbol{\toctoura}$ -- 
the window of vulnerability between two successive \ra instances performed by a device, 
during which the state of the software is unknown and potentially compromised without detection; 
colored orange in Figure \ref{fig: toctou window}(a). \toctoura can be exploited by 
transient malware that: (1) infects a device, (2) remains active for a 
while, and (3) erases itself and restores the device software to its ``good'' state, as 
long as (1)-(3) occur between two successive \ra instances.

\noindent $\boldsymbol{\toctousa}$ -- the inter-device \toctou window,
i.e., the time variance between the earliest and the latest \ra performed across networked
devices. colored red in Figure \ref{fig: toctou window}(a).
Consider a situation where, the verifier performs network attestation. Device-A receives an 
attestation request, performs its attestation at time $t_0$ and replies to the verifier.
At time $t_1>t_0$, the verifier receives device-A's attestation report, checks it, and decides that 
device-A is benign. However, device-A is compromised at time $t_2>t_1$. Meanwhile, due to network delay, device-B performs its attestation at $t_3>t_2$ and replies.
The verifier receives device-B's attestation report at $t_4>t_3$ and 
(erroneously) concludes that both devices are now benign.

\noindent \textbf{Synchronized Attestation} -- An important requirement for \sa is 
that all attestation reports should accurately reflect current system state.
If devices are attested at different times, the verifier cannot determine 
if the network as a whole is (or was) in a secure state, even if all individual 
\ra reports reflect the benign state. This undermines trust in current attestation methods, 
motivating the need for more synchronized network-wide attestation.

\noindent {\bf Performance Overhead:}
Attesting the entire software state of a 
device is computationally expensive. For safety-critical IoT devices, 
minimizing time spent on non-safety-critical tasks (e.g., \ra) is crucial to 
maintain responsiveness and real-time performance. Even a lightweight \ra, 
which is typically based on a device computing a relatively fast Message Authentication Code (MAC) 
(usually implemented as a keyed hash) requires doing so over the entire application program memory.
This introduces a non-negligible delay which is a function of memory size. For example, a TI MSP430 
microcontroller unit (MCU) running at $8$MHz takes $\approx450$ms to compute an
SHA2-256 HMAC over $4$KB of program memory \cite{vrased}. 
This delay is significant for real-time or safety-critical systems with 
tight timing constraints. 

\noindent {\bf Energy Overhead:}
Execution of \ra consumes
energy on battery-powered or energy-harvesting IoT devices. This is particularly 
problematic for devices deployed in remote or inaccessible locations where 
battery replacement is difficult or infeasible. Reducing power consumption 
is therefore both beneficial and important. 

\noindent {\bf Unreliable Communication:}
Malware-infected devices can 
subvert the attestation process by dropping or modifying attestation 
requests and replies. Prior techniques do not adequately address this 
problem.

\noindent To this end, we construct \system: \systemtext. It 
offers two protocol variants: \trapsrtc\ -- for devices equipped with  
real-time clocks (RTCs), and \trapsnortc\ -- for devices without such clocks. 
\system is designed to work with low-end IoT devices that have a small set 
of security features, based on \rata \cite{rata} or \casu \cite{casu} \ra techniques
which were originally developed for a single-device \ra setting. \system pairs
these \ra techniques with \garota\cite{garota} -- another recent technique that constructs
a minimal {\bf active} Root-of-Trust (\rot) for low-end devices and guarantees
operation even if a device is fully malware-compromised. Specifically, \system\
uses NetTCB and TimerTCB of \garota which ensure, respectively: (1) timely
sending and receiving messages, and (2) starting attestation on time, 
with no interference from any other software.

\noindent Contributions of this work are:
\begin{compactenum}[(1)]
\item \textbf{Reduced \toctou Window}: \system employs time synchronization 
(using RTCs or a depth-based mechanism) to ensure nearly simultaneous 
attestation across all devices in the network, substantially reducing the \toctousa window. 
The \toctoura window is mitigated by the use of \casu or \rata security features.
Figure \ref{fig: toctou window}(b) shows decreased \toctou windows by \system.
\item \textbf{Efficient and resilient \ra}: \system combines a few \rot constructions 
to minimize \ra-induced performance overhead and power consumption for individual 
devices, while guaranteeing timely \ra execution by isolating it from any potential 
malware interference.
\item \textbf{Open-Source Implementation}: \system's practicality and 
cost-effectiveness is confirmed via a fully functional prototype 
on a popular low-end IoT device platform -- TI MSP430 micro-controller,
using FPGA \cite{TRAPSAnonOpenSource}.
\end{compactenum}

\section{Background}\label{sec:bg}
\subsection{Targeted Devices}
We focus on resource-constrained devices using low-end MCUs, such as Atmel AVR ATmega and TI MSP430, 
which are low-power single-core platforms with limited memory. These devices have $8$-bit or $16$-bit 
CPUs, $1$-$16$MHz clock frequencies, and typically $\leq~64$KB of addressable memory. Data memory
(DMEM) ranges from $4$ to $16$KB, while the rest is program memory (PMEM). Software executes in-place
from PMEM. It runs on ``bare metal'',  with no memory management for virtual memory or isolation.

A representative architecture for targeted devices includes a CPU core, DMA controller, and 
interrupt controller connected via a bus to memory regions: ROM, PMEM, DMEM, and peripheral 
memory. ROM holds the bootloader and immutable software. Device software resides in PMEM, 
and DMEM is used for the stack and heap. The device may incorporate both internal peripherals (timers) and external peripherals (sensors, actuators).

\subsection{Remote Attestation (\ra)}
As mentioned earlier, \ra is used for malware detection on a remote device. It is typically
achieved via a challenge-response protocol that enables a trusted entity called a verifier (\vrf) 
to remotely verify the software state of an untrusted remote device (\prv):
\begin{compactenum}  
  \item \vrf sends an \ra request with a challenge (\chal) to \prv.
  \item \prv generates an unforgeable attestation report, i.e., an authenticated integrity check 
  over PMEM, including the software, and \chal, and sends it to \vrf.
  \item \vrf verifies the report to determine whether \prv is in a valid state.
\end{compactenum}
The report includes either a Message Authentication Code (MAC) or a signature, depending 
on the type of cryptography used. In the former case, \prv and \vrf must share a unique 
secret key -- \key, while in the latter, \key is a unique private key of \prv. In either case, 
\key must be stored securely and be accessible only to the trusted 
attestation code on \prv.  

A large body of research 
\cite{vrased,noorman2013sancus,smart,nunes2020apex,rata,casu,jakkamsetti2023caveat,
aldoseri2023symbolic,wang2023ari,evtyushkin2014iso,ghaeini2019patt,sehatbakhsh2019emma,
feng2021scalable} explored \ra for low-end devices. Prior work can be split into {\bf passive} and 
{\bf active} techniques.
The former only {\bf detects} compromise and offers no guarantee of the device responding to an 
\ra request. Whereas, the latter either {\bf prevents} compromise and/or 
guarantees small security-critical tasks (e.g., an \ra response).

\subsection{Network Attestation (\sa)}
Unlike single-device \ra, which involves one \vrf and one \prv, \sa deals with a potentially
large number (network, group, or swarm) of \prv-s. This opens new challenges.
First, na\"{\i}ve adoption of single-\prv{} \ra techniques is inefficient and even impractical.
Also, \sa needs to take into account topology discovery, key management, and routing. This can 
be further complicated by mobility (i.e., dynamic topology) and device heterogeneity.
Moreover, \toctousa (inter-device \toctou) emerges as a new problem.

\subsection{Building Blocks} 
\noindent
{\bf $\boldsymbol{\rata}$ \cite{rata}} is a passive Root-of-Trust (\rot) architecture that 
mitigates \toctoura with minimal additional hardware. \rata securely logs the last PMEM 
modification time to a protected memory region called Latest Modification Time (LMT), 
which can not be modified by any software. 
\prv's attestation report securely reflects the integrity of its software state indirectly 
through the LMT. This approach is based on the principle that any modification to the 
software in PMEM would necessitate an update to the LMT. Thus, by attesting the LMT, \rata 
effectively attests the software state without needing to read the entire PMEM contents. 
This minimizes \ra computational overheads of \prv by attesting only 
a fixed-size ($32$-byte) LMT (plus the \vrf's challenge of roughly the same size), instead 
of attesting its entire software in PMEM.

\noindent
{\bf $\boldsymbol{\casu}$ \cite{casu}} is an active \rot architecture 
that provides run-time software immutability and authenticated software updates. 
It defends against code injection (into PMEM) and data execution attacks by preventing 
(1) unauthorized modification of PMEM and (2) code execution from DMEM.
\casu monitors several CPU hardware signals (e.g., program counter, write-enable bit, and 
destination memory address) and triggers a reset if any violation is detected.
The only means to modify PMEM is via secure update.
\casu inherently prevents \toctoura since PMEM cannot be overwritten by malware.

\noindent
{\bf $\boldsymbol{\garota}$ \cite{garota}} is another active architecture which 
guarantees execution of trusted and safety-critical tasks. 
These tasks are triggered based on arbitrary events captured by hardware peripherals 
(e.g., timers, GPIO ports, and network interfaces), even if malware is present on the device.
\garota provides two hardware properties: ``guaranteed triggering’’ and 
``re-triggering on failure’’.
The former ensures that a particular event of interest always triggers execution of 
\garota TCB tasks,
while the latter ensures that if TCB execution is interrupted for any reason (e.g., attempts to 
violate execution integrity), the device resets, and the TCB task is guaranteed to be 
executed first after the boot. \garota has 3 flavors:  TimerTCB, NetTCB, and GPIO-TCB,. 
In this paper, we are interested in the first two:
(1) {\bf TimerTCB} -- A real-time system where a predefined safety-critical task is 
guaranteed to execute periodically, and
(2) {\bf NetTCB} -- A trusted component that guarantees to process commands 
received over the network, thus preventing malware on the MCU from intercepting 
and/or discarding commands destined for the \rot.

\subsection{Hash Chains for Authentication \label{chains}}
Hash chains provide a secure, scalable, and efficient means of authentication,
originally proposed by Lamport \cite{lamport1981}. 
Over the last 40+ years, they have been used in numerous settings where one party (signer/sender)
needs inexpensive (though limited or metered) authentication to a multitude of receivers.

An $m$-link hash chain is constructed by repeatedly applying a cryptographic hash function 
$H$, starting with the initial secret value $x_0$, such that:
\[\Scale[0.85]
{H(x_0) = x_1, H(x_1) = x_2, ... , H(x_{m-1})=x_m, \;\;\mbox{i.e.,}\;\;
x_m = H^m(x_0)}
\]
To set up the operation of an $m$-link hash chain, signer (sender) retains
$x_0$ ({\bf root}) and shares final $x_m$ ({\bf anchor}) 
with all receivers. Given a value $x_i$ where $(0\leq~i\leq~m)$, it is computationally 
infeasible to compute $x_{i-1}$ or any previous value $x_k$ for $k < i$.
Conversely, calculating $x_{i+1}$ or any subsequent value $x_j$ where $j > i$ is straightforward;
$x_j$ can be computed by repeatedly applying the hash function $H()$ to $x_i$ (j-i times). 

For the first authentication round, signer reveals $x_{m-1}$ and all receivers can easily
authenticate it by comparing $H(x_{m-1})\stackrel{?}{=}x_m$. In the second round, 
the signer reveals $x_{m-2}$, and so on. This continues until the hash chain is exhausted, 
at which point a new hash chain is generated and shared. See Section \ref{renewal} 
for the use of hash chains in \system.

Suppose that a receiver is de-synchronized: it currently has $x_i$ and, instead
of the expected $x_{i-1}$, it next receives authenticator $x_j$ where $j<(i-1)$. This means that 
this receiver missed $i-j-1$ successive authenticators: $x_{i-1},...,x_{j+1}$.
Nonetheless, a receiver can quickly re-synchronize by authenticating $x_j$ via
computing $H^{(i-j)}(x_j)$ and checking if it matches $x_i$.

\section{Design Overview \label{design}} 
\subsection{System Model}
\noindent {\bf Network:}
We assume a single verifier (\vrf) and a network of multiple low-end embedded devices as 
\prv-s. \vrf is assumed to be trusted and sufficiently powerful. 
The network is assumed to be: (1) connected, i.e., there is always a path between \vrf and any of  
\prv-s, and (2) quasi-static during attestation, i.e., its topology can change as long as the changes 
do not influence the path of message propagation. \system is network-agnostic and can be 
realized over any popular medium (e.g., WiFi, Bluetooth, Cellular, Zigbee, Matter).

\noindent {\bf $\boldsymbol{\ra}$ Architecture in $\boldsymbol{\prv}$:}
All \prv-s must support \rata or \casu architecture: in a given deployment, either all support
the former or all support the latter, i.e., no mixing.\footnote{This is not a hard requirement,
meaning that a mix of \rata and \casu devices would work as well;
however, it makes the presentation simpler.}
As mentioned in Section \ref{sec:bg}, an attestation token in 
\rata is computed as a keyed hash over a small fixed-size input.

In contrast, \casu prevents any PMEM modification (except via secure update), 
thus obviating the entire need for \ra. However, \casu does not offer \prv liveness.
Note that, in any secure \ra technique, an attestation token returned by \prv 
provides both attestation and \prv liveness.
The latter is important for detecting whether \prv is operational, i.e., not powered off, 
destroyed/damaged, or physically removed. 
To this end, \casu supports a ``secure heartbeat'' feature, whereby 
\vrf periodically issues a random challenge and \prv simply computes (and returns) a keyed 
hash over that challenge. This costs about the same as attestation token computation in \rata. 
We discuss various use-cases of \rata and \casu in Section \ref{subsec:rata-casu-comparison}.

\noindent {\bf Network Interface in $\boldsymbol{\prv}$:}
The primary network interface of each \prv is placed within \system's Trusted Computing Base (TCB). 
This ensures that \system protocol messages are handled with the highest 
priority, even in the presence of malware or run-time attacks.
\system uses two special attestation-specific packet types: request and report. 
Normal software outside TCB is prevented from sending or receiving these packet types;
this is achieved by inspecting each incoming/outgoing packet header
in order to prevent tampering with, and forgery of, \system messages. Furthermore, 
\system packets are always handled with higher priority than other tasks. 
This approach is based on NetTCB of \garota \cite{garota}.
Besides, we adopt TimerTCB from \garota to guarantee (nearly) synchronized attestation start times.

With these security features, \rata-enabled \prv-s are safeguarded against
full compromise and malware-based disruption of the attestation process.
The benefit is more subtle in the case of \casu: although \casu guarantees no malware,
software running on \casu-enabled \prv-s can still be susceptible to control-flow attacks, 
which would prevent, or delay, receiving of \vrf attestation requests and 
generation of secure heartbeats. The above features ensure that this does not occur.

\noindent {\bf $\boldsymbol{\prv}$ TCB:}
\system TCB includes both hardware and software components, i.e., akin to \rata and \casu, 
\system is a {\bf hybrid} architecture. In addition to the trusted software of either \rata or \casu, 
\system software includes TimerTCB, NetTCB, and \sa logic described in Section \ref{sec:protocol}.
The primary network interface (NetTCB) is shared between the \system software  
and other non-TCB software. Incoming messages cause an interrupt via NetTCB.
\system software prioritizes \system protocol messages. It forwards other incoming messages to the intended
application (outside TCB) and outgoing messages to the destination.
TCB hardware components are:
\begin{compactitem}
    \item NetTCB -- Network interface for \system messages
    \item TimerTCB -- Timer used for simultaneous attestation
    \item DMEM segment reserved for running \system software 
    \item Part of ROM reserved for \system software, key shared with \vrf, and hash chain data
\end{compactitem}
\subsection{Adversary Model \label{sec: adversary}}
In line with other network attestation (\sa) techniques, \system considers software-only 
remote network attacks. We assume an adversary (\sadv) that can inject malware and exercise 
full control over a compromised \prv, except for its TCB. \sadv can manipulate any non-TCB
peripherals and external components, such as Direct Memory Access (DMA), sensors, actuators, 
and other (non-primary) network interfaces. Also, \sadv has comprehensive knowledge 
of software (i.e., non-\system software) running on \prv,
including its memory vulnerabilities. Thus, it can launch run-time (e.g., control-flow) attacks.

We also consider a network-based \sadv represented by a malicious (non-\system) entity in the
\prv network. Consequently, all packets exchanged between \vrf and \prv-s can 
be manipulated by \sadv: based on the Dolev-Yao model \cite{dolevYao}, \sadv can eavesdrop on, 
drop, delay, replay, modify, or generate any number of messages. 

\noindent {\bf DoS Attacks:}
\system prevents DoS attacks that attempt to ``brick'' \prv-s via malware,
or control-flow attacks. However, DoS attacks that jam the network or attempt to inundate 
specific \prv's network interfaces are out of scope. 
For countermeasures, we refer to well-known techniques, 
such as \cite{muraleedharan2006jamming,zhijun2020low, mamdouh2018securing}.

\noindent {\bf Physical Attacks:}
\system does not offer protection against physical attacks, both 
invasive (e.g., via hardware faults and reprogramming through debuggers) and non-invasive (e.g., 
extracting secrets via side-channels). Such attacks can be mitigated, at considerable cost, via
well-known tamper-resistance methods \cite{obermaier2018past,ravi2004tamper}.

\subsection{Protocol Elements \label{elems}}
As mentioned in Section \ref{intro}, we construct two \system variants, 
based on the availability of a real-time clock (RTC) on \prv-s:
\begin{compactenum}[(1)]
    \item \textbf{$\boldsymbol{\trapsrtc}$:} Each \prv has an RTC. In an attestation request, \vrf includes 
    the exact time when all \prv-s should perform attestation. 
    \item \textbf{$\boldsymbol{\trapsnortc}$:} \prv-s do not have RTC-s. In an attestation request, 
    \vrf provides the height of the spanning tree, composed of all \prv-s. Each \prv 
    estimates the time to perform attestation using spanning tree height and its own secure timer.
\end{compactenum}

\noindent \trapsrtc is designed for an ideal best-case scenario where each \prv is assumed to 
have a synchronized RTC. \toctousa window is completely removed in \trapsrtc.
On the other hand, \trapsnortc is intended for a more realistic scenario where each \prv has a timer.
Although \trapsnortc can not offer precisely synchronized attestations on \prv-s, it still significantly reduces \toctousa. Section \ref{subsec:security-analysis} provides further details.

\noindent{\bf $\boldsymbol{\toctousa}$ resilience}
Due to the availability of RTC in \trapsrtc and the spanning tree's height in \trapsnortc, 
all \prv-s perform attestation almost simultaneously. Figure \ref{fig: toctou window}(b) 
shows the eliminated \toctou window in \trapsrtc.

\noindent {\bf Attestation Regions:}
Unlike prior \sa schemes which perform attestation over the 
entire PMEM, \system is built on top of either \rata or \casu, which enables \prv to 
compute a MAC over a short fixed size including:
(1) \lmt\ and \vrf's challenge, in \rata,  or (2) merely \vrf's challenge, in \casu.
Section \ref{sec:protocol} provides details about other parameters included in the MAC computation.

\noindent {\bf Authentication of Attestation Requests:}
Most prior work in network (or swarm) attestation
does not take into account authentication of attestation requests. While this may or may
not be an issue in a single \prv{} \ra setting\footnote{\vrf authentication in a single \prv 
setting is thoroughly discussed in \cite{Brasser-DAC16}.}, it certainly becomes a concern in \sa.
If requests are not authenticated, \sadv can readily mount a DoS attack whereby \sadv floods all 
\prv-s with bogus requests, each of which causes {\bf all} \prv-s to perform attestation and 
generate numerous useless replies.

This issue is deceptively simple. The na\"ive approach to address the problem is for \vrf
(which already shares a unique symmetric key with each \prv) to send 
an individual attestation request to every \prv, authenticated with each shared key.
This is unscalable for obvious reasons. 

Another intuitive approach is to assume that every \prv knows \vrf's
public key and \vrf simply signs each attestation request with a timestamp. Despite scaling well, this 
approach opens the door for a simple DoS attack whereby \sadv floods the network with 
attestation requests with fake signatures, forcing all \prv-s to verify them, and due to failed
verification, discard the requests. This incurs heavy collective computational 
overhead on the entire network.

Yet another trivial method is to assume a separate group key (shared among \vrf and all \prv-s) that is
used exclusively for authenticating \vrf-issued attestation requests. This is quite efficient
since a simple MAC (realized as a keyed hash) would suffice. However, a key shared among a potentially
large number of \prv-s raises the risk of its eventual compromise, which would have 
unpleasant consequences. 
Also, managing the group and key revocation becomes increasingly complex as the network grows.

\system uses hash chains to authenticate attestation requests. Hash chains, as described in Section 
\ref{chains}, are well-known constructs used in numerous similar settings where symmetric keys 
are unscalable and traditional public key signatures are too expensive.
They provide forward security and efficient verification, while offering relatively simple
key management. Although hash chains suffer from some fragility in terms of synchronization 
and timing requirements, these issues are more palatable than those that stem from managing 
large numbers of shared keys.

\section{\system Protocols} \label{sec:protocol}
This section describes two protocol variants.
The notation used in the rest of the paper is summarized in Table \ref{table:notation}.

\noindent{\bf Assumptions:}
As mentioned above, we assume that each \prv shares a unique symmetric key (\key) with \vrf. 
Also, throughout a single attestation instance, \vrf is assumed to be within the broadcast range 
of at least one \prv, and the entire \prv network must remain connected during this time.
Furthermore, all \prv-s have a parameter (\maxdelay) that denotes the maximum attestation report (\Attrep)
propagation delay in the network. In the absolute worst case of a line topology, it can
be set as: $\maxdelay=n*t_{report}$, where $n$ is the number of \prv-s and $t_{report}$ is the \Attrep\ propagation delay.
We also assume that the attestation request (\Attreq) propagation delay ($t_{request}$) 
and the \Attreq\ verification time ($t_{hash}$) are known to all \prv-s.
\maxdelay\ is needed to limit the time when 
each \prv forwards other \prv-s' attestation results towards \vrf.
For the sake of simplicity, we assume that no new attestation requests 
are issued while one is being served.

\noindent {\it NOTE:} As mentioned at the end of Section \ref{elems}, the use of hash 
chains for \vrf authentication is optional; a separate group key shared between \vrf and all \prv-s 
could be used instead, albeit with the risk of its possible leak.

\subsection{\trapsrtc: RTC-Based \sa Technique}\label{sec:one}
Commodity RTCs, such as MCP7940MT-I/SM \cite{rtc}, are now readily available for 
under $\$0.60$ per unit. This affordability marks a significant shift from the past, 
when real-time security features were often too costly for IoT devices. This motivates 
our design of an \sa protocol for devices with RTCs.
We begin by presenting this simple variant of the core ideas of \system.
An alternative variant without the RTC requirement is described in Section \ref{sec:two}.
\trapsrtc pseudo-code is shown in Algorithms \ref{alg:rtc_prv} and \ref{alg:rtc_vrf}.

\begin{figure}
    \centering
    \includegraphics[height=1.2in,width=0.7\columnwidth]{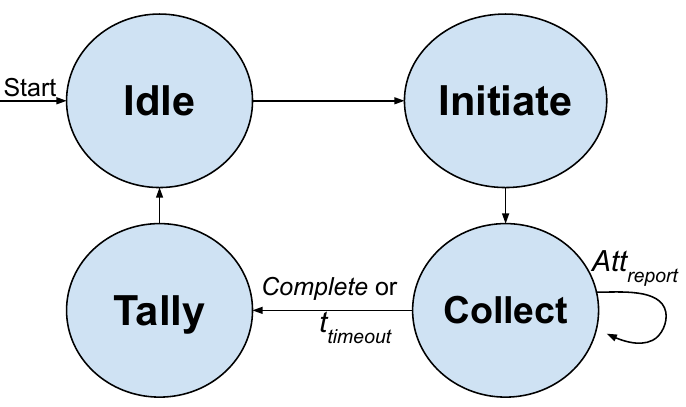}
    \vspace{-.5em}
    \caption{\vrf\ State Machine}
    \label{fig:vrfFSM}
    \vspace{-1.5em}
\end{figure}
\begin{compactitem}[]
    \item {\bf \underline{\bf V1.} Idle:} \vrf waits for an external signal to begin an attestation instance. 
    When it occurs, \vrf transitions to \textbf{Initiate}.
    \item {\bf \underline{\bf V2.} Initiate:} \vrf assigns \attesttime{}, 
    as described in Section \ref{section:timeout}, which accounts for request 
    propagation and network height. (\attesttime{} is computed by each \prv in \trapsnortc.)     
    It then initializes 
    \textit{Attest}=\textit{Fail}=$\emptyset$, and \textit{NoRep}=\{\devid-s of all \prv-s\}. 
    Next, \vrf sets \snd\ to \vrf, composes \Attreq, and broadcasts it. (Recall the assumption 
    that at least one \prv must be within broadcast range of \vrf at this time.) 
    It then sets a local timer to \timeout{},
    as detailed in Section \ref{section:timeout}, which factors into network size and delays, 
    and then transitions to \textbf{Collect}.
    \item {\bf \underline{\bf V3.} Collect:} \vrf waits for \Attrep-s. 
    Upon receipt of an \Attrep, \vrf first checks that \hash\ contained in \Attrep\ matches that in the 
    currently pending \Attreq; otherwise, it is discarded. Next, \vrf validates \Attrep\ by 
    looking up the corresponding \key shared with \prv (identified by \devid) and recomputing 
    the MAC. If MAC validation fails, \Attrep\ is discarded. Otherwise:
    \begin{compactitem}[]
        \item {\it \underline{V3.1}} (\casu): \devid\ is moved from \textit{NoRep} to \textit{Attest}.
        \item {\it \underline{V3.2}} (\rata): \vrf maintains the last valid \lmt\ for each \prv. 
    When processing an \Attrep\ from a given \prv, \vrf compares received \lmt$'$\ with the stored 
    \lmt\ for that \prv. A mismatch signifies failed attestation and
    \devid\ is added to \textit{Fail}. Otherwise, it is added to \textit{Attest}. 
    In either case, \devid\ is removed from \textit{NoRep}.
        \item {\it \underline{V3.3}} If $\textit{NoRep}=\emptyset$, \vrf transitions to \textbf{Tally}.
        \item {\it \underline{V3.4}} If \timeout\ timer expires, \vrf transitions to \textbf{Tally}.
    \end{compactitem}
    \item {\bf \underline{\bf V4.} Tally:} \vrf outputs \textit{Attest}, \textit{Fail}, and 
    \textit{NoRep}. It then returns to \textbf{Idle}.
\end{compactitem}

\begin{algorithm}[hbt!]
\footnotesize
\caption{Pseudo-code of $\boldsymbol{\trapsrtc}$ for \prv}\label{alg:rtc_prv}
    \begin{algorithmic}[1]
    \While{True}
        \State \textit{m} = RECEIVE()
        \If {TYPE(\textit{m}) == \textit{``req''}}
            \State [\snd, \hashind, \hash, \attesttime] $\gets$ DECOMPOSE(\textit{m})
            \If {$\curhashind <= \hashind$}
                \State CONTINUE() 
            \EndIf
            \If {GET\_TIME() $>=$ \attesttime}
                \State CONTINUE() 
            \EndIf
            \If {\textit{H}$^{(\curhashind-\hashind)}(\hash)$ $\neq$ \curhash}
                \State CONTINUE()
            \EndIf
             \State \parent $\gets$ \snd; \curhashind $\gets$ \hashind; \curhash $\gets$ \hash; \textit{attestTime} $\gets$ \attesttime;
             \State BROADCAST(\textit{``req''}, \devid, \curhashind, \curhash, \attesttime)
             \State \textit{CurTime} $\gets$ \textit{GET\_TIME()} \Comment{Get current time from RTC}
             \While {\textit{CurTime} $<$ \attesttime} \Comment{non-busy-waiting}
             \State \textit{CurTime} $\gets$ GET\_TIME()
             \EndWhile
             \State \attesttime$'$ $\gets$ \textit{CurTime}
             \State \Authrep $\gets$ MAC(\key, \parent, \attesttime$'$, \hash, \{\lmt\})
             \State \Attrep $\gets$ \textit{``rep''}, \devid, \parent, \attesttime$'$,  \hash, \{\lmt\}, \Authrep
             \State UNICAST(\parent, \Attrep)
             \State SET\_TIMER(\height*t$_{report}$)
        \EndIf
         \If {TYPE(\textit{m}) == \textit{``rep''}}
                 \If {\hash == GET\_\hash(\textit{m})}
                     \State UNICAST(\parent, \textit{m})
                 \EndIf
        \EndIf
    \EndWhile
    \end{algorithmic}
\end{algorithm}

\begin{algorithm}[hbt!]
\footnotesize
\caption{Pseudo-code of $\boldsymbol{\trapsrtc}$ for \vrf}\label{alg:rtc_vrf}
    \begin{algorithmic}[1]
    \While{True}
        \State \textit{type} $\gets$ REQUEST\_TYPE() 
        \State \hashind $\gets$ GET\_HASH\_IND()
        \State \hash $\gets$ GET\_HASH(\hashind)
        \State \attesttime $\gets$ \netheight*(t$_{request}$+t$_{hash}$)+t$_{slack}$+\textit{GET\_TIME()}
        \State \Attreq $\gets$ \textit{``req'', vrf}, \hashind, \hash, \attesttime
        \State \textit{Attest} $\gets \emptyset$; \textit{Fail} $\gets \emptyset$; \textit{NoRep} $\gets$ \{\devid-s of all \prv-s\}
        \State BROADCAST(\Attreq)
        \State T $\gets$ GET\_TIME()
        \While{\attesttime $<$ T $<$ \timeout}
             \State \Attrep $\gets$ RECEIVE()
             \State [\devid, \parent, \attesttime, \hash, \lmt, \Authrep] $\gets$ DECOMPOSE(\textit{\textit{m}})
             \State \lmt$'$ $\gets$ LMT\_LIST(\devid) \Comment{\casu skips \#13, \#15, \#17, \#18}
             \If{(\hash \ == \curhash) \ AND 
             (MAC(\key, \parent, \attesttime, \hash, \{\lmt\}) == \Authrep)}
                 \If {\lmt == \lmt$'$} 
                     \State \textit{Attest} $\gets$ \textit{Attest} $\cup$ \devid
                 \Else
                     \State \textit{Fail} $\gets$ \textit{Fail} $\cup$ \devid
                 \EndIf
             \State \textit{NoRep} $\gets$ \textit{NoRep} $\backslash$ \devid
             \EndIf
        \EndWhile
        \State OUTPUT(\textit{Attest}, \textit{Fail}, \textit{NoRep})
    \EndWhile
    \end{algorithmic}
\end{algorithm} 

\trapsrtc\ has two message types:

\noindent \textbf{Attestation Request ($\boldsymbol{\Attreq}$):} Generated by \vrf, 
it contains: \hash, \hashind, and \attesttime, which are used to authenticate \Attreq. Note that \hash\ 
is used as a challenge for this \sa. $\boldsymbol{\Attreq}$ also includes
the packet type field: \textit{``req''} and the identifier \snd\ of either \vrf 
that originated it (for the first hop), or a \prv that forwards it (for subsequent hops).
\snd\ is used by each receiving \prv to learn its parent in the spanning tree.

\noindent \textbf{Attestation Report ($\boldsymbol{\Attrep}$):} Generated by each \prv, this 
message carries the attestation report. It contains an authentication token (\Authrep), which provides message integrity. 
\lmt$'$\ is included in the calculation of \Authrep\ and in \Attrep\ only for \rata-enabled \prv-s. 
Similar to $\boldsymbol{\Attreq}$, $\boldsymbol{\Attrep}$
also includes the packet type field: \textit{``rep''}  and the identifier of \prv that generated 
this report. Also, \Attrep\ includes \hash\ that was received in \Attreq\ and the actual time 
(\attesttime$'$) when attestation is performed.

\begin{table}[ht]
    \vspace{.7em}
    \footnotesize
    \begin{tabularx}{\linewidth}{|c|X|}
        \hline
        \rowcolor{gray!20}
        {\bf Notation} & \multicolumn{1}{c|}{\bf Meaning} \\            
        \thickhline
        $\devid$ & Identifier of responding $\prv$ \\
        \hline
        $\parent$ & Identifier of responding $\prv$'s parent \\
        \hline
        $\snd$ & Identifier of the sending device \\
        \hline
        $H()$ & Cryptographic hash function (e.g., SHA2-256) used in hash chain computation \\
        \hline 
        $H^s(x)$ & Denotes $s>1$ repeated applications of $H()$ starting with initial input $x$ \\
        \hline        
        $\hash$ & Hash value sent by $\vrf$ that authenticates it to all $\prv$-s; 
            it also serves as the challenge for this $\sa$ instance \\
        \hline
        $\curhash$ & Current hash value stored by $\prv$ \\
        \hline
        $\hashind$ & Index of $\hash$ sent by $\vrf$  \\
        \hline
        $\curhashind$ & Index of $\curhash$ stored by $\prv$ \\
        \hline
        $\netheight$ & Network spanning tree height  \\
        \hline
        $\height$ & Height of $\prv$ in the spanning tree \\
        \hline        
        $\lmt$ & Last Modification Time (of PMEM), only used in $\rata$, stored on \vrf \\        
        \hline
        $\lmt'$ & Last Modification Time (of PMEM), only used in $\rata$, stored on \prv \\        
        \hline
        \key & Shared key between \prv and \vrf, securely stored on \prv and restricted to its trusted attestation code \\
        \hline
        $\Attreq$ & \parbox[c]{1\linewidth}{\vspace{2pt}\raggedright Attestation request message ($\vrf \rightarrow \prv$): \\
        \phantom{}[\textit{"req"}, \snd, \hash, \hashind, \attesttime]}\vspace{1pt}\\
\hline
        $\Attrep$ & \parbox[c]{1\linewidth}{\vspace{2pt}\raggedright Attestation report message ($\vrf \leftarrow \prv$): \\
        \phantom{}\tiny{[\textit{"rep"}, \devid, \parent, \attesttime$'$, \hash, \{\lmt\}, \Authrep]}}\vspace{1pt}\\
\hline
        $\Authrep$ & \parbox[c]{1\linewidth}{\vspace{2pt}\raggedright 
           Authentication of attestation report in $\Attrep$: \\ 
           \tiny{MAC(\key, \parent, \attesttime$'$, \hash, \{\lmt$'$\})}}\vspace{1pt}\\
        \hline
        t$_{request}$ & propagation delay of $\Attreq$\\
        \hline
        t$_{report}$ & propagation delay of $\Attrep$\\
        \hline
        t$_{hash}$ & Computation time for $\Attreq$ verification \\
        \hline        
        t$_{MAC}$ & Computation time for MAC generation \\
        \hline
        t$_{slack}$ & Additional slack time \\
        \hline
        $\maxdelay$ & Max delay to receive an $\Attrep$ from a descendant $\prv$ \\
        \hline
        $\attesttime$ & Time to begin attestation, set by $\vrf$ \\
        \hline        
        $\attesttime'$ & Time when a given $\prv$ actually performed attestation \\
        \hline
        $\timeout$ & $\vrf$'s timeout for receiving all attestation replies \\
        \hline
    \end{tabularx} 
    \caption{Notation Summary} 
    \vspace{-3em}
    \label{table:notation}
\end{table}

\begin{figure}
    \centering
    \includegraphics[height=1.5in,width=0.7\columnwidth]{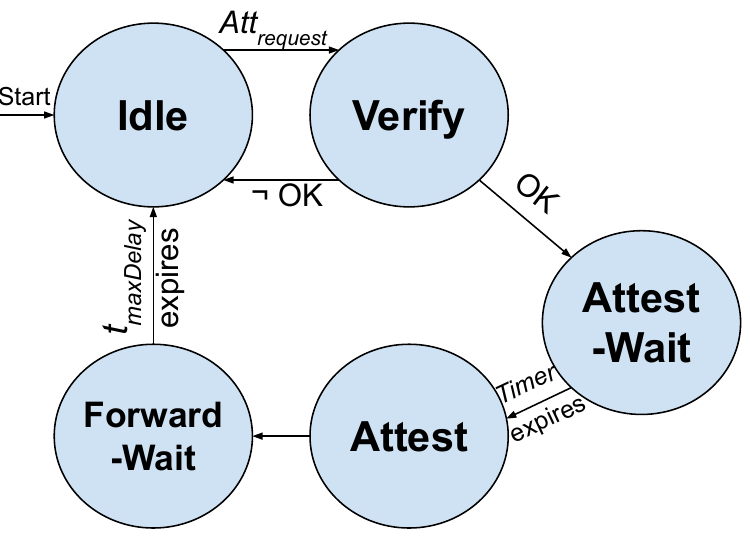}
    \vspace{-1em}
    \caption{\prv\ State Machine} 
    \label{fig:prvFSM}
\end{figure}

We now describe \prv operation as a state machine with five states, as shown in 
Figure \ref{fig:prvFSM}: {\it Idle, Verify, Attest-Wait, Attest,}  and {\it Forward-Wait}.

\begin{compactitem}[]
    \item {\bf \underline{\bf P1.} Idle:} \prv runs normally. Upon receiving an \Attreq, it 
    proceeds to \textbf{Verify}. Any \Attrep\ received in this state is discarded.
    \item {\bf \underline{\bf P2.} Verify:} 
    \begin{compactitem}[]
        \item \underline{\it P2.1:} \prv checks if $\curhashind>\hashind$ and $\attesttime>T$,
    where $T$ is its current RTC value. If either check fails, it discards \Attreq\ and 
    returns to {\bf Idle}. 
        \item \underline{\it P2.2:} \prv computes and checks whether $H^s(\hash) \stackrel{?}{=} 
    \curhash$, where $s=\curhashind-\hashind$.\footnote{Recall that it is possible for $s>1$,
    (as discussed at the end of Section \ref{chains}) meaning that \prv became de-synchronized.
    Also, \hashind\ is decremented by one in every \ra instance.}
    If not, it discards \Attreq\ and returns to {\bf Idle}.
    (Note that a \prv might receive duplicate \Attreq-s from multiple neighbors; it simply discards them.)
    \item \underline{\it P2.3:} \prv\ replaces: \curhash\ with \hash, and \curhashind\ with \hashind. 
    Then, \prv stores \snd\ as \parent, sets \snd\ field of received \Attreq\ to its \devid, and 
    broadcasts modified \Attreq.
    \item \underline{\it P2.4:} \prv sets (using its RTC) a secure timer (TimerTCB) to \attesttime\ 
    and transitions to \textbf{Attest-Wait}.
    \end{compactitem} 
     \item {\bf \underline{\bf P3.} Attest-Wait:} \prv runs normally while the timer is ticking. 
     If any \Attreq\ is received in this state, it is discarded. 
    \item {\bf \underline{\bf P4.} Attest:} When the timer matches \attesttime, \prv sets {\it \attesttime$'$} 
    to the current RTC value, computes \Authrep{}, and composes 
    \Attrep\ as defined above. It then uni-casts \Attrep\ to \parent{}, sets the timer 
    to \maxdelay, and transitions to \textbf{Forward-Wait}.
%
    \item {\bf \underline{\bf P5.} Forward-Wait:} \prv runs normally while the timer is ticking. If \prv 
    receives an \Attrep, it checks whether the report's \hash\ matches that previously received 
    in \textbf{Verify}. If not, it is discarded. Otherwise, \prv uni-casts received \Attrep\  
    to its parent and remains in \textbf{Forward-Wait}. When the timer matches \maxdelay, \prv transitions 
    to \textbf{Idle}. Note that any \Attreq \ received while in this state 
    is discarded. 
\end{compactitem}

Whereas, as shown in Figure \ref{fig:vrfFSM}, \vrf's state machine has four states: 
{\it Idle, Initiate, Collect,} and {\it Tally}.

\subsection{\trapsnortc: Clockless \sa Technique}\label{sec:two}

Despite its relatively low cost, an RTC might still not be viable for some IoT deployments. 
This leads us to construct a \system variant without RTCs.
Pseudo-code for \trapsnortc is shown in Algorithms \ref{alg:nortc_prv} and \ref{alg:nortc_vrf}.

\begin{algorithm}[hbt!]
\footnotesize
 \caption{Pseudo-code of $\boldsymbol{\trapsnortc}$ for \prv}\label{alg:nortc_prv}
     \begin{algorithmic}[1]
     \While{ True }
         \State \textit{m} = RECEIVE()
         \If {TYPE(\textit{m}) == ``req''}
             \State [\snd, \hashind, \hash, \height, \netheight, \attesttime] $\gets$ DECOMPOSE(\textit{m})
             \If {$\curhashind <= \hashind$}
                 \State CONTINUE() 
             \EndIf
             
             \If {\textit{H}$^{(\curhashind-\hashind)}$(\hash) $\neq$ \curhash}
                 \State CONTINUE()
             \EndIf
             \State \parent $\gets$ \snd; \curhashind $\gets$ \hashind; \curhash $\gets$ \hash; \textit{attestTime} $\gets$ \attesttime;
             \State BROADCAST(\textit{``req''}, \devid, \curhashind, \curhash, \height + 1, \netheight, \attesttime)
        
             \State \textit{timer} $\gets$ \textit{startTimer()} \Comment{start a timer}
             \State \textit{attestWait} $\gets$ (\netheight-\height)*(\textit{t$_{request}$+\textit{t$_{hash}$})}
        
             \While {\textit{timer} $<$ \textit{attestWait}} \Comment{non-busy-waiting}
                 \State WAIT()
             \EndWhile
             \state \attesttime$'$ $\gets$ \textit{timer}
             \State \Authrep $\gets$ MAC(\key, \parent, \attesttime, \hash, {\lmt})
             \State \Attrep $\gets$ \textit{``rep''}, \devid, \parent, \attesttime$'$, \hash, {\lmt}, \Authrep
             \State UNICAST(\parent, \Attrep)
             \State SET\_TIMER(\height*\textit{t$_{report}$})
         \EndIf
         \If {TYPE(\textit{m}) == \textit{``rep''}}
                 \If {\hash == GET\_\hash(\textit{m})}
                     \State UNICAST(\parent, \textit{m})
                \EndIf
        \EndIf
    \EndWhile
     \end{algorithmic}
 \end{algorithm}

 \begin{algorithm}[hbt!]
\footnotesize
 \caption{Pseudo-code of $\boldsymbol{\trapsnortc}$ for \vrf}\label{alg:nortc_vrf}
     \begin{algorithmic}[1]
     \While{True}
         \State \textit{type} $\gets$ REQUEST\_TYPE() 
         \State \hashind $\gets$ GET\_HASH\_IND()
         \State \hash $\gets$ GET\_HASH(\hashind)
         \State \attesttime $\gets$ \textit{\netheight}*(t$_{request}$+$t_{hash})+t_{slack}$+\textit{GET\_TIME()}
         \State \netheight $\gets$ GET\_NET\_height()
    
         \State \Attreq $\gets$ \textit{``req''}, \vrf, \hashind, \hash, 0, \netheight, \attesttime
         \State Attest $\gets \emptyset$; Fail $\gets \emptyset$; NoRep $\gets$ \{\devid-s of all \prv-s\}
         \State BROADCAST(\textit{InitID}, \Attreq)
         \State T $\gets$ GET\_TIME()
        \While{T $<$ \timeout}
             \State \Attrep $\gets$ RECEIVE()
             \State [\devid, \parent, \attesttime, \hash, \{\lmt\}, \Authrep] $\gets$ DECOMPOSE(\textit{\textit{m}})
             \State \lmt$'$ $\gets$ LMT\_LIST(\devid) \Comment{\casu skips \#14, \#16, \#18, \#19}
             \If{(\hash \ == \curhash) \ AND  (MAC(\key, \parent, \attesttime, \hash, \{\lmt\}) == \Authrep)}
                 \If {\lmt == \lmt$'$}
                     \State \textit{Attest} $\gets$ \textit{Attest} $\cup$ \{\devid\}
                 \Else
                     \State \textit{Fail} $\gets$ \textit{Fail} $\cup$ \{\devid\}
                 \EndIf
             \State \textit{NoRep} $\gets$ \textit{NoRep} $\backslash$ \{\devid\}
             \EndIf
        \EndWhile
        \State OUTPUT(\textit{Attest}, \textit{Fail}, \textit{NoRep})
     \EndWhile
     \end{algorithmic}
 \end{algorithm} 

There are still just two message types, \Attreq\ and \Attrep, 
of which only \Attreq\ differs from \trapsrtc:

\noindent \textbf{Attestation Request ($\boldsymbol{\Attreq}$):} Generated by \vrf, \Attreq\ includes two extra fields: \height\ and \netheight\ 
which represent the height of the sender (\vrf or \prv) and the spanning tree height of the network,
respectively. \height\ is essentially a hop counter during the propagation of \Attreq\ 
throughout the network. It is initialized to 0 by \vrf and incremented by each forwarding \prv.

\prv's state machine has five states, three of which are identical to those in \trapsrtc.
Only {\underline{Verify}} and {\underline{Attest}} differ, as follows:
\begin{compactitem}[]
    \item {\bf \underline{\bf P2.} Verify:} 
    \begin{compactitem}[]
        \item \underline{\it P2.1:}  \prv checks whether $\curhashind>\hashind$.    
        If this check fails, it discards \Attreq\ and returns to {\bf Idle}.
        \item \underline{\it P2.2:} \prv 
        computes and checks whether $H^s(\hash) \stackrel{?}{=} \curhash$, where 
        $s=\curhashind-\hashind$. If not, it discards \Attreq\ and returns to {\bf Idle}.
        Duplicate \Attreq-s from multiple neighbors are also discarded. 
        \item  \underline{\it P2.3:} \prv replaces: \curhash\ with \hash, and \curhashind\ with \hashind. Then, \prv stores \snd\ as \parent, 
        sets \snd\ field to its \devid, increments \height{}, and broadcasts modified \Attreq. 
        \item \underline{\it P2.4:} \prv sets a secure timer (TimerTCB) to: 
        \begin{equation}
            \textit{attestWait} = (\netheight-\height)*({t_{request}} + {t_{hash}})
        \end{equation}
        and transitions to \textbf{Attest-Wait}.
    \end{compactitem}
    \item {\bf \underline{\bf P4.} Attest:} Identical to \trapsrtc, except that \prv sets {\it \attesttime$'$} 
    to its current secure timer value, to be later validated by \vrf.
    The degree of reduction of \toctousa depends on the accuracy and functionality of \prv's secure timer.
    Also, the propagation delay from \vrf to each \prv affects \toctousa.
    This is discussed in more detail in Section \ref{subsec:security-analysis}.
\end{compactitem}
Note that \attesttime\ in \trapsnortc is a timer value (increases from 0),
unlike that in \trapsrtc, which represents the current time.
\vrf's state machine has four states identical to that of \trapsrtc.

\noindent {\bf Protocol Trade-offs:} \system's two variants address distinct deployment 
constraints. \trapsrtc uses RTCs to synchronize attestation timing globally via precise 
timestamps (\attesttime), thus minimizing \toctousa with marginal hardware costs.
Whereas, \trapsnortc eliminates RTC dependencies by deriving attestation timing from the 
network topology (\netheight), thus sacrificing \toctousa precision for broader applicability.
These synchronization implications are addressed in Section \ref{dis:precise_sync}.

\subsection{Renewing Hash Chains \label{renewal}}
As typical for any technique utilizing hash chains, the issue of chain depletion must
be addressed. An $m$-link hash chain is depleted after $m$ authentication instances
($m$ \sa instances in our context). To address this issue and ensure 
long-term operation, we need a mechanism for refreshing the hash chain. 

Recall the well-known Lamport hash chain construct from Section \ref{chains}. 
Suppose that the current hash chain of length $m$ being used is ${\mathcal X}$:
$$
H(x_0) = x_1, H(x_1) = x_2, ... , H(x_{m-1})=x_m
$$
Suppose that we have already used up $m-2$ links of the chain for all \prv-s. 
This means that only two links in the chain remain, and the entire chain will be depleted when \vrf reveals $x_1$ and then
$x_0$ in the next two \sa instances. Knowing this, \vrf wants all \prv-s 
to switch over to a new hash chain ${\mathcal X'}$:
$$
H(x'_0) = x'_1, H(x'_1) = x'_2, ... , H(x'_{m-1})=x'_m
$$
To do so, it includes in the next \Attreq\ (\Attreq$_{_{m-1}}$) two extra values/fields:
\begin{center}
\noindent 
\Attreq$_{_{m-1}}$ = [\textit{``req''}, \snd, \hash=$x_1$, \hashind=$1$, \\
\attesttime, {\bf NewChain=$\boldsymbol{x'_m}$, Auth=$\boldsymbol{MAC(x_0,x'_m)}$} ]
\end{center}
These two new fields convey the anchor of the new hash chain {\bf NewChain} and its authenticator {\bf Auth} 
computed as a MAC over NewChain using, as a key, still-unreleased next link in the current chain -- $x_0$. 
Upon receiving such an \Attreq, in addition to the usual \Attreq \ processing, 
a \prv stores NewChain and Auth. Obviously, at this time, a \prv has no way to 
verify Auth since it does not yet know $x_0$. A \prv continues to process this \Attreq,
as detailed earlier.

However, at this stage, each \prv{} maintains a current hash ${\mathcal X}$, where 
$\curhashind=1$ and $\curhash=x_1$. A \prv waits for the next \sa instance, wherein 
\Attreq$_{_m}$ should convey $x_0$. Upon receiving \Attreq$_{_m}$, a \prv obtains 
$x_0'$, which may differ from the original $x_0$ if it was modified by \sadv in transit.
As part of its normal processing, a \prv first verifies that $H(x_0')=\curhash=x_1$. 
A \prv recomputes Auth$'$ using the newly received $x_0'$ and its stored NewChain value. 
If Auth$'$ matches the previously stored Auth, a \prv completes the switchover 
to the chain ${\mathcal X'}$ by setting $\curhashind=m$ and $\curhash=x'_m$. 

This simple renewal technique is secure, lightweight, and trivial to implement. 
However, two factors contribute to its fragility. 

\noindent {\bf Timing:}
It must hold that the time difference between \vrf sending \Attreq$_{_{m-1}}$
and \Attreq$_{_{m}}$ is sufficiently long to avoid forgeries of NewChain and Auth. However, even 
when the time difference is reasonably long, \sadv can delay the delivery of \Attreq$_{_{m-1}}$ 
to one or more targeted \prv-s. If \Attreq$_{_{m}}$ is sent by \vrf
when at least one \prv has not yet received \Attreq$_{_{m-1}}$, \sadv can learn $x_0$ from \Attreq$_{_{m}}$. 
It can then change the NewChain field in \Attreq$_{_{m-1}}$  from $x'_m$ to $y_m$, and Auth field --
from $MAC(x_0,x'_m)$ to $MAC(x_0,y_m)$, where $y_m$ is the anchor of \sadv-selected hash chain.

This issue is not unique to the present technique. It is indeed quite similar to the timing requirement
in the well-known TESLA protocol for secure multicast and its many variants \cite{perrig-tesla}.
TESLA also uses the delayed key disclosure mechanism and makes reasonable assumptions about 
timing.\footnote{See Section 2.2 in IETF RFC 4082: \url{https://www.ietf.org/rfc/rfc4082.txt)}.}
The timing issue can be further mitigated if \vrf switches the chain in 
\Attreq$_{_{m}}$ only if it has received legitimate responses from all \prv-s upon
sending \Attreq$_{_{m-1}}$. 

\noindent {\bf DoS on $\boldsymbol{\prv}$-s:}
Upon observing \Attreq$_{_{m-1}}$, \sadv (present in the network) can modify 
NewChain and/or Auth fields. Each \prv would then duly store these two values. Once the subsequent 
\Attreq$_{_m}$ arrives in the next \sa instance, each \prv would fail to verify stored NewChain and Auth, 
thus ending up being unable to process any further \Attreq-s. Although there is no full-blown fix for 
this problem, one way to side-step it is for \vrf to begin switching to the new hash chain prior 
to a few links being left in the old chain, i.e., when \curhash \ $=(m-k)$ for some reasonably small $k$.
In this case, Auth $={MAC(x_{m-k-1},x'_m)}$, which can be verified in the successive attestation, 
\Attreq$_{_{m-k-1}}$. Then, \vrf can decide to switch to ${\mathcal X'}$ when it receives valid 
\Attrep-s from all \prv-s, indicating that all have the identical NewChain, ${x'_m}$.

\subsection{Timeouts}\label{section:timeout}
The overall attestation timeout (on \vrf) is set as follows:
\begin{equation}\label{eq:t-overall}
    \timeout = n * (t_{request}+t_{hash} + t_{report}) + t_{MAC} + t_{slack}
\end{equation}
where $n$ is the total number of \prv-s in the network.
\vrf sets the attestation time in \trapsrtc as follows:
\begin{equation}\label{eq:t-attestation}    
    \attesttime = \netheight * (t_{request}+t_{hash})+t_{slack}+t_{current}
\end{equation}
where t$_{current}$ is \vrf's current time.

\attesttime\ must be large enough for every \prv to receive \Attreq\
before the actual attestation begins.
Note that an inflated \attesttime\ does not influence \toctousa; it only incurs \vrf's waiting time.
In the worst case (line topology), the total request propagation time would be: $n*(t_{request}+t_{hash})$.
Once all devices receive the request, they perform attestation at (ideally) the 
same time \attesttime, taking \ $t_{MAC}$. Finally, \Attrep-s from all \prv-s 
need to be returned to \vrf, which takes at most $n*t_{report}$ in the worst case. 
Note that t$_{report}$ may differ from t$_{request}$ \ due to network congestion caused by 
simultaneous response transmissions from all \prv-s. An additional tolerance
value t$_{slack}$ helps account for unexpected delays.
\section{Implementation \label{impl}}
\system is prototyped atop openMSP430\cite{openMSP430}, an open-source implementation of TI MSP430 MCU, 
written in Verilog HDL. OpenMSP430 can execute software generated by any MSP430 toolchain 
\cite{msp430-gcc} with near-cycle accuracy. We extended both \rata and \casu architectures to 
support \system. In this implementation, \prv and \vrf are connected via UART.

\subsection{\system Software}
Using the native msp430-gcc toolchain, \system software on \prv is compiled to generate software images 
compatible with the memory layout of the modified openMSP430. \system software, responsible for processing 
\system protocol messages and generating attestation responses, is housed in ROM.
NetTCB is triggered whenever a \system protocol message is received; this is determined by the cleartext 
message type in the header.

Also, TimerTCB is triggered to start attestation whenever the timer expires in the {\bf Attest-Wait} state.
For cryptographic operations we use a formally verified cryptographic library, HACL* \cite{hacl}.
It provides high-assurance implementations of essential cryptographic primitives, such as hash 
functions and MAC-s. SHA2-256 and HMAC are used for hash and MAC, respectively.
Both \rata and \casu implement their respective cryptographic operations using HACL*.

To emulate \vrf, we developed a Python application with $\approx~200$ lines of code, 
as described in Sections \ref{sec:one} and \ref{sec:two}. The application runs on an
Ubuntu 20.04 LTS laptop with an Intel i5-11400 processor @$2.6$GHZ with $16$GB of RAM.

\subsection{\system Hardware}

As mentioned earlier, \prv-s in \system can adopt either \casu or \rata architecture,
possibly equipped with different system resources (e.g., CPU clock, 
memory, peripherals). We refer to \casu-based \prv-s as $\boldsymbol{\trapscasu}$ and 
\rata-based \prv-s as $\boldsymbol{\trapsrata}$. We implemented and evaluated both 
as part of the proof-of-concept.

The design is synthesized using Xilinx Vivado 2023.1, a popular logic synthesis tool. 
It generates the hardware implementation for the FPGA platform. The synthesized design 
is then deployed on a Basys3 Artix-7 FPGA board for prototyping and evaluating hardware design.

\begin{figure}
    \captionsetup{justification = centering}
    \centering
    \includegraphics[height=1.3in,width=0.8\columnwidth]{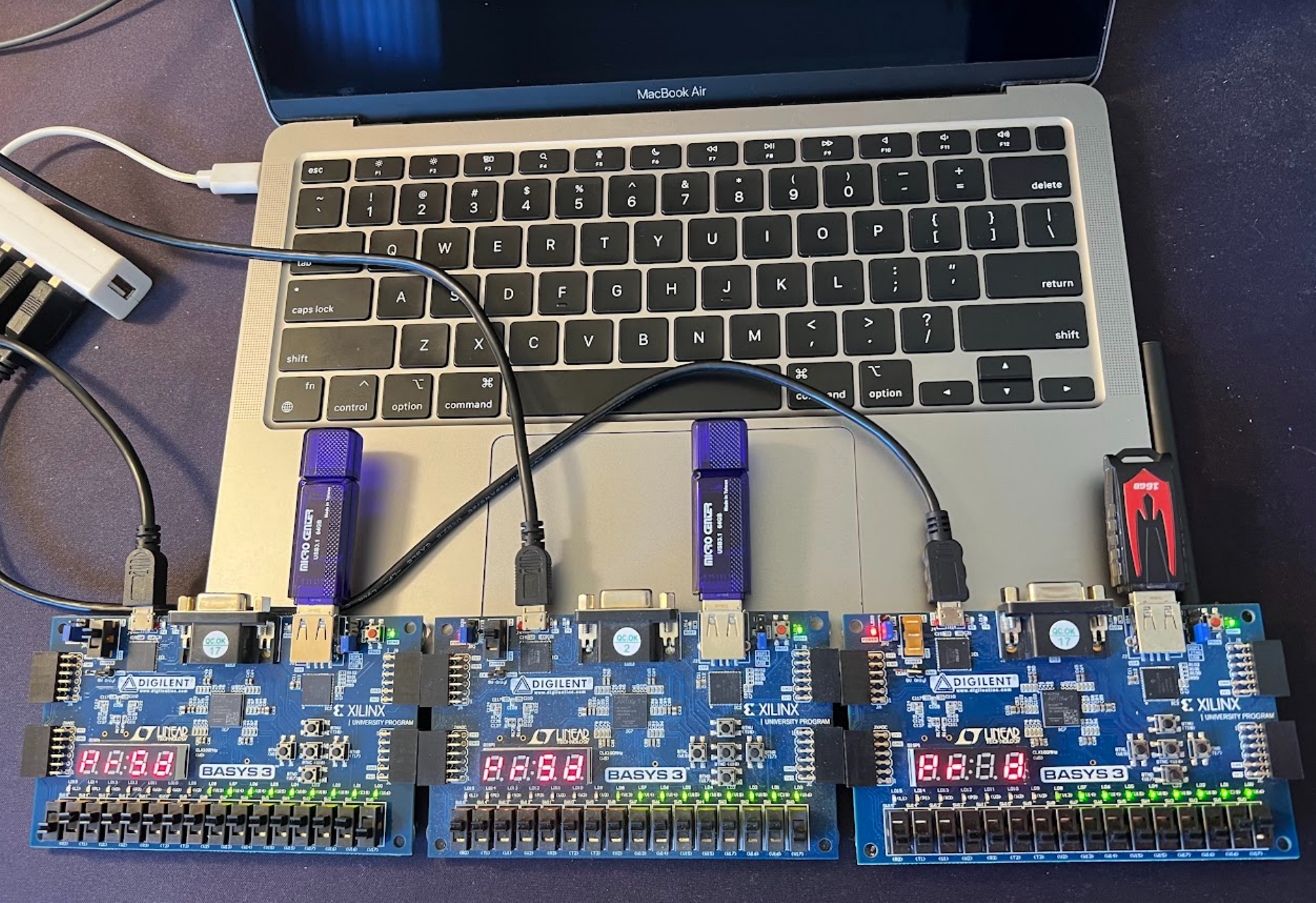}
    \caption{\system Proof-Of-Concept with Three \prv-s}  
    \vspace{-1.5em}
    \label{fig: PoC}
\end{figure}

Figure \ref{fig: PoC} shows a proof-of-concept implementation of \system. In it,
three \prv-s (implemented on Basys3 FPGA boards) are connected to \vrf. For the sake of simplicity, 
\prv-s are deployed using a star topology for signal routing. All three \prv-s in Figure \ref{fig: PoC} are \trapscasu 
devices. However, we also implemented \system with \trapsrata devices for performance evaluation.

\section{Evaluation \label{eval}}

\subsection{Security Analysis} \label{subsec:security-analysis}

\noindent {\bf Network-based $\boldsymbol{\sadv}$:}
This adversary (\sadv) is a malicious (not a \system{} \prv) physical network entity, e.g., a non-compliant IoT device or a computer. 

\Attreq\ in \trapsrtc includes: \textit{``req''}, \snd, \hash, \hashind, \attesttime, while \Attreq\ in \trapsnortc also includes \height{} and \netheight. \prv authenticates each \Attreq{} by verifying \hashind, \attesttime, and checking if $H^s(\hash) = \curhash$, where $s=\curhashind-\hashind$. The \hash{} is known only to \vrf, and recovering it from \curhash{} is computationally infeasible, so \sadv cannot forge \hash. However, \sadv can modify other fields (such as \attesttime, \height, and \netheight) affecting \prv's attestation time. Nonetheless, this is later detected by \vrf, as \attesttime$'$ is included in \Authrep{} within each \prv's \Attrep.

\sadv can also alter the \snd{} field in \Attreq{}, supplying an incorrect \parent{} to \prv. This may obstruct valid \Attrep-s from benign \prv-s. However, \vrf will notice the absence of \Attrep{} from affected \prv-s.

\Attrep{} includes: \textit{``rep''}, \devid, \parent, \attesttime$'$, \hash, \{\lmt$'$\}, and \Authrep, with authenticity and integrity ensured by \Authrep, computed as:\\ MAC(\key,\parent,\attesttime$'$,\hash,\{\lmt$'$\}). Manipulation of \devid\ is detectable by \vrf since \devid is used to retrieve the corresponding key.

\sadv can forge an \Attrep{} only if: (1) \sadv forges \Authrep{} without knowing \key, which is infeasible with a secure MAC function, or (2) \sadv learns \key and constructs an authentic \Authrep{}, which is infeasible since \key is in the TCB and is only accessible to \system software.

\noindent {\bf Malware-based $\boldsymbol{\sadv}$:}
\system remains secure despite malware presence on any number of \prv-s due to: (1) \prv's TCB ensuring \key secrecy, (2) NetTCB enforcing receiving and forwarding of \Attreq\ and \Attrep\, (3) TimerTCB ensuring timely \Attrep\ generation, and (4) \prv's TCB blocking non-TCB software from sending \Attrep\ or \Attreq\ messages. These measures prevent DoS attacks from \prv-resident malware.

\noindent {\bf $\boldsymbol{\toctousa}$ \& $\boldsymbol{\toctoura}$:}
\trapsrtc eliminates \toctousa as long as the RTC is accurately synchronized with the \vrf. Meanwhile, minimizing \toctousa in \trapsnortc depends on: (a) the secure timer, and (b) propagation delay from \vrf to each \prv. Two scenarios relating to the former could increase \toctousa in \trapsnortc: (a-1) \sadv tampering with a \prv's secure timer, or (a-2) timer drift due to physical imperfections, disrupting the attestation schedule.

To address (a-1), \trapsnortc uses TimerTCB to: (1) prioritize the timer's Interrupt Service Routine (ISR) for timely attestation, and (2) protect timer configurations from unauthorized changes. Although (a-2) can't be fully addressed, \trapsnortc significantly reduces \toctousa compared to unsynchronized schemes. For example, with a propagation delay $t_{request}=1$ms and \netheight$=10,000$, \prv-s wait for up to $10$s. A timer drift of $100$ppm results in a $1$ms drift, reducing \toctousa from $10,000$ms to $1$ms in \trapsnortc.

 Recall that \trapsnortc assumes identical network propagation delays. However, in reality, variations may occur due to congestion or connectivity changes. For instance, with \netheight$=10,000$ and $t_{request}=1$ms, if the delay between $\prv_i$ and $\prv_j$ is $1.5$ms, $\prv_i$ starts attestation $0.5$ms earlier than its descendants. To minimize this, (b), $t_{request}$ should average all network propagation delays. Note that \netheight\ over-estimation by \vrf doesn't affect \toctousa; it only delays attestation start on \prv-s.

\begin{table}[!t]
    \centering
    \small
    \begin{tabularx}{\linewidth}{|c|X|}
        \hline
        \rowcolor{gray!20}
        {\bf Notation} & \multicolumn{1}{c|}{\bf Description} \\            
        \thickhline
        $\pc$	     & 	Program Counter pointing to the current instruction being executed \\
        \hline
        $\wen$ 	 &	1-bit signal that represents whether MCU core is writing to memory \\
        \hline
        $\daddr$ 	 &	Memory address being accessed by MCU core \\
        \hline
        $\dmaen$   &	1-bit signal that represents whether DMA is active \\
        \hline
        $\dmaaddr$ &	Memory address being accessed by DMA \\
        \hline
        $\reset$	 &	Signal that reboots the MCU when it is set to logic `1'\\
        \hline
        $\tcr$	 &	Trusted code region, a fixed ROM region storing $\trapscasu$ software \\
        \hline        
        $\er$	     &	Executable region, a memory region where authorized (normal) software is stored \\
        \hline
        $\ep$	     &	Executable pointer, a fixed memory region storing current $\er$ boundary \\    
        \hline
        $\ivtr$	 &	Reserved memory region for the MCU's interrupt vector table \\
        \hline
        $\casumem$ &  Memory region protected by $\trapscasu$ hardware, including $\er$, $\ep$, and $\ivtr$ \\
        \hline
        $\gie$     &  Global interrupt enable, 1-bit signal that represents whether interrupts are globally enabled \\
        \hline
        $\irq$     &  1-bit signal that represents if an interrupt occurs \\
        \hline
        $\irqcfg$  &  Set of registers in DMEM used to configure of interrupts, e.g., timer deadline and UART baudrate\\
        \hline
        $\isrt$    &  Timer interrupt service routine, privileged software that controls a timer interrupt: \newline
                        $\isrt = [\isrtmin, \isrtmax]$ \\    
        \hline
        $\isru$    &  UART interrupt service routine, privileged software that handles a UART interrupt: \newline
                        $\isru = [\isrumin, \isrumax]$ \\
        \hline        
    \end{tabularx}
    \caption{Notation Summary}
\label{table:notations}
\end{table}

\begin{figure*}[t]
    \begin{mdframed}[userdefinedwidth=0.95\textwidth]
        $\bullet\ $ {\it Security Properties Stemming from} $\boldsymbol{\casu}$\\
        \textbf{- \small Software Immutability in PMEM:}\\        
        \begin{equation}\label{eq:hwprop_wp}
            \textbf{G}: \{\text{modMem}(\casumem) \land (\pc \notin \tcr) \rightarrow \reset\}
        \end{equation}
        \textbf{- \small Unauthorized Software Execution Prevention:}
        \begin{equation}\label{eq:hwprop_usep}
            \textbf{G}: \{(\pc \notin \er) \land (\pc \notin \tcr) \rightarrow \reset\}
        \end{equation} \\
        $\bullet\ $ {\it Security Properties Stemming from} $\boldsymbol{\garota}$\\
        \textbf{- \small IRQ Configuration Protection:}
        \begin{equation}\label{eq:irqcfg-protection}
            \textbf{G}: \{[\neg(\pc \in \tcr) \land \wen \land (\daddr \in \irqcfg)] \lor [\dmaen \land (\dmaaddr \in \irqcfg)] \rightarrow \reset\}
        \end{equation}
        \textbf{- \small Timer ISR Execution Atomicity:}        
        \begin{equation}\label{eq:timer-isr-protection1}
            \textbf{G}: \{\neg \reset \land \neg (\pc \in \isrt) \land (\textbf{X}(\pc) \in \isrt) \rightarrow \textbf{X}(\pc) = \isrtmin \lor \textbf{X}(\reset) \}
        \end{equation}
        \begin{equation}\label{eq:timer-isr-protection2}
            \textbf{G}: \{\neg \reset \land (\pc \in \isrt) \land \neg (\textbf{X}(\pc) \in \isrt) \rightarrow \pc = \isrtmax \lor \textbf{X}(\reset) \}            
        \end{equation}
        \begin{equation}\label{eq:timer-isr-protection3}
            \textbf{G}: \{(\pc \in \isrt) \land (\irq \lor \dmaen) \rightarrow \reset \}
        \end{equation}
        \textbf{- \small UART ISR Execution Atomicity:}        
        \begin{equation}\label{eq:uart-isr-protection1}
            \textbf{G}: \{\neg \reset \land \neg (\pc \in \isru) \land (\textbf{X}(\pc) \in \isru) \rightarrow \textbf{X}(\pc) = \isrumin \lor \textbf{X}(\reset) \}
        \end{equation}
        \begin{equation}\label{eq:uart-isr-protection2}
            \textbf{G}: \{\neg \reset \land (\pc \in \isru) \land \neg (\textbf{X}(\pc) \in \isru) \rightarrow \pc = \isrumax \lor \textbf{X}(\reset) \}            
        \end{equation}
        \begin{equation}\label{eq:uart-isr-protection3}
            \textbf{G}: \{(\pc \in \isru) \land (\irq \lor \dmaen) \rightarrow \reset \}
        \end{equation}
        \textbf{- \small Interrupt Disablement Protection:}
        \begin{equation}\label{eq:irq-disablement-protection}
            \textbf{G}: \{\neg \reset \land \gie \land \neg \textbf{X}(\gie) \rightarrow (\textbf{X}(\pc) \in (\isrt \lor \isru)) \lor \textbf{X}(\reset)\}
        \end{equation}        
    \end{mdframed}
    \caption{\trapscasu Hardware Security Properties}\label{fig:hwprop}
\end{figure*}

\noindent{\bf Formal Verification of $\boldsymbol{\trapscasu}$:}
We formally specify \trapscasu with \trapsnortc security goals using Linear Temporal Logic (LTL).
Formal verification plays a crucial role by showing that \trapscasu adheres to well-specified goals.
It assures that it does not exhibit any unintended behavior, especially in corner cases, 
rarely encountered conditions and/or execution paths, that humans tend to overlook.
By employing computer-aided tools, we define and validate LTL rules that govern \trapscasu operation.
The use of LTL enables precisely capturing temporal dependencies and expected behavior 
over time, ensuring that \trapscasu meets stringent security standards.
Table \ref{table:notations} describes the notation used in this section.

We use regular propositional logic, such as conjunction $\land$, disjunction $\lor$, 
negation $\neg$, and implication $\rightarrow$. A few other temporal quantifiers are used as well:
\begin{compactitem}
    \item $\textbf{X} \Phi$ (ne$\textbf{X}$t) -- holds if $\Phi$=true at the next system state.
    \item $\textbf{F} \Phi$ ($\textbf{F}$uture) -- holds if there is a future state when $\Phi$=true.
    \item $\textbf{G} \Phi$ ($\textbf{G}$lobally) -- holds if for all future states $\Phi$=true.
\end{compactitem}
Figure \ref{fig:hwprop} formally describes \trapscasu hardware security properties using 
propositional logic and temporal quantifiers. Recall that \trapscasu is based on \casu 
combined with \garota. All such properties must hold at all times to achieve \trapscasu's
security goals.

LTL \ref{eq:hwprop_wp} states that any modifications to \casumem, including \er, \ep, and \ivtr, 
trigger a reset when \trapscasu software is not running. \er is a region in PMEM, where normal 
device software resides, while \ep is a fixed region in PMEM that points to \er.
Upon a secure update, \ep is updated to point to the new verified software version.
\ivtr also resides in PMEM and contains the ISR addresses.
As stated in LTL \ref{eq:hwprop_usep}, the MCU cannot execute any code outside \er 
or \trapscasu code in read-only memory (ROM).

LTL \ref{eq:irqcfg-protection} ensures that, if the timer or the UART peripheral configurations
are modified by any software (other than the timer or UART ISR-s), a reset is triggered.
LTL \ref{eq:timer-isr-protection1}-\ref{eq:timer-isr-protection3} specify atomic operation of 
timer ISR, LTL \ref{eq:timer-isr-protection1} and LTL \ref{eq:timer-isr-protection2} guarantee 
that \isrtmin and \isrtmax are the only legal entry and exit points, respectively.
Also, LTL \ref{eq:timer-isr-protection3} states that DMA and other interrupts must remain 
inactive while timer ISR executes. Similarly, LTL 
\ref{eq:uart-isr-protection1}-\ref{eq:uart-isr-protection3} enforce UART ISR atomicity.
Finally, LTL \ref{eq:irq-disablement-protection} guarantees that \gie can be disabled 
only if the timer or UART ISR-s are running. Any violations result in a device reset.

Note that we slightly modified \casu and \garota to realize \trapscasu:
\begin{compactenum}
    \item \trapscasu employs both TimerTCB and NetTCB, while \garota uses them individually in each case.
    \item {\it Trusted PMEM Updates} rule from \garota is integrated to Equation \ref{eq:hwprop_wp}.
    \item \garota's {\it Re-Trigger on Failure} property is not viable 
    since \trapscasu cannot retain a consistent timer value upon a failure (e.g., a reset) in \trapsnortc. 
\end{compactenum}

To verify the above LTL rules, we convert the Verilog code described at the Register Transfer Level 
(RTL) to Symbolic Model Verifier (SMV) \cite{mcmillan1993smv} using Verilog2SMV \cite{irfan2016verilog2smv}.
The SMV output is in turn fed to the NuSMV \cite{cimatti2002nusmv} model-checker for specified LTL 
rule validation. NuSMV works by checking LTL specifications against the system finite-state machine 
for all reachable states. This comprehensive approach ensures that \trapscasu's security goals 
are thoroughly validated, offering robust assurance against potential vulnerabilities.
See \cite{TRAPSAnonOpenSource} for further proof details.

\subsection{Hardware Overhead} \label{subsec:hw_overhead}
%
Recall that underlying hardware RoT for \prv-s in \system is either \casu or \rata with 
additional hardware support from \garota.
Table \ref{table: hardware overhead} compares the hardware overhead of \trapscasu and \trapsrata
implementations with the baseline openMSP430, \casu, and \rata architectures.
\trapscasu implementation requires $0.46$\% more Look-Up Tables (LUTs) and $0.55$\% more registers over 
\casu.  Also, \trapsrata implementation needs $0.05$\% LUTs and $0.69$\% registers over \rata.
Numbers of additional LUTs and registers are under $15$, implying minimal overheads incurred
by NetTCB and TimerTCB.

\begin{figure}
	\centering\captionsetup{justification = centering}
	\subfloat[Additional LUTs (\%)]
	{\includegraphics[width=0.5\columnwidth]{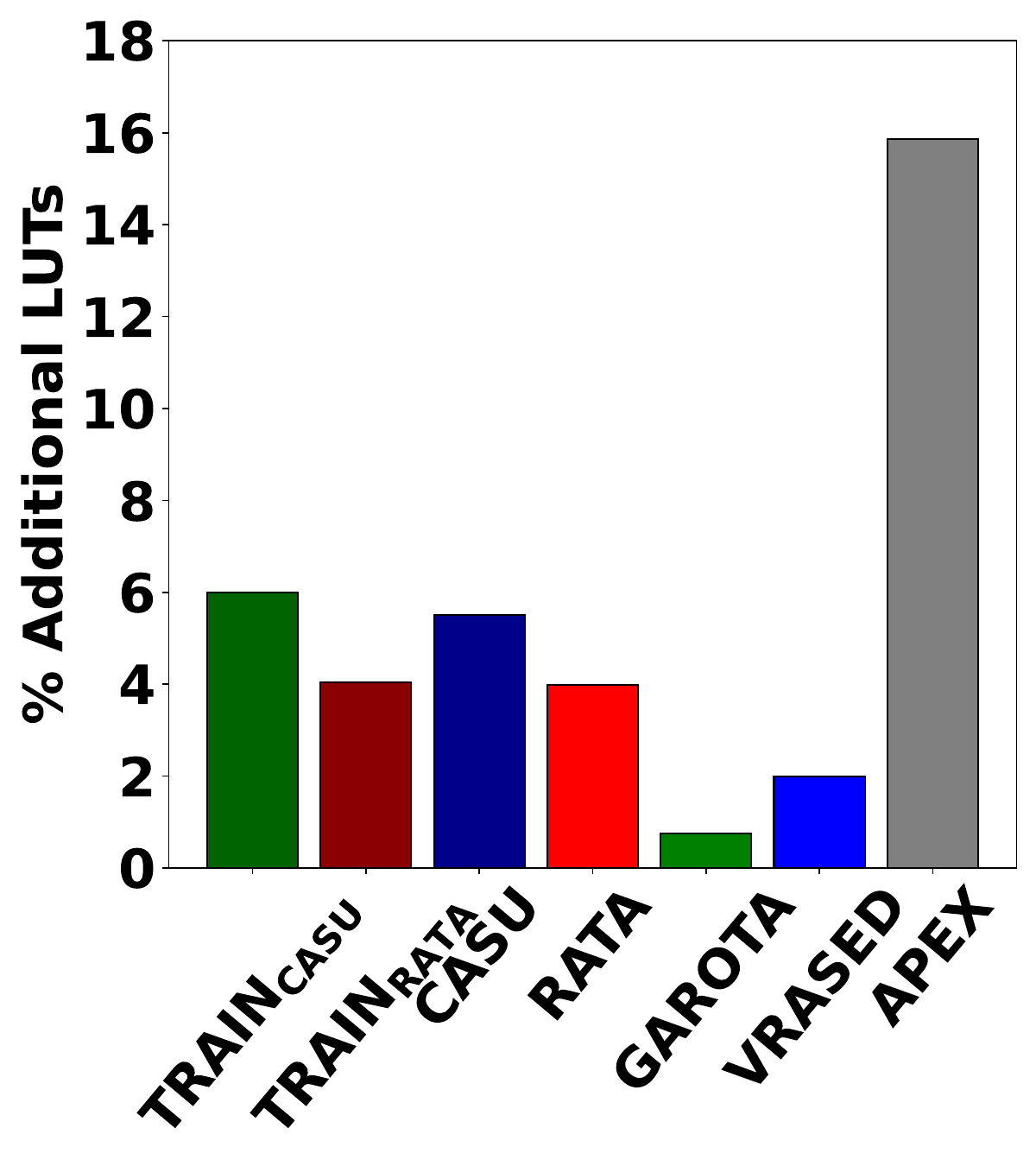}}
	\subfloat[Additional Reg-s (\%)]
	{\includegraphics[width=0.5\columnwidth]{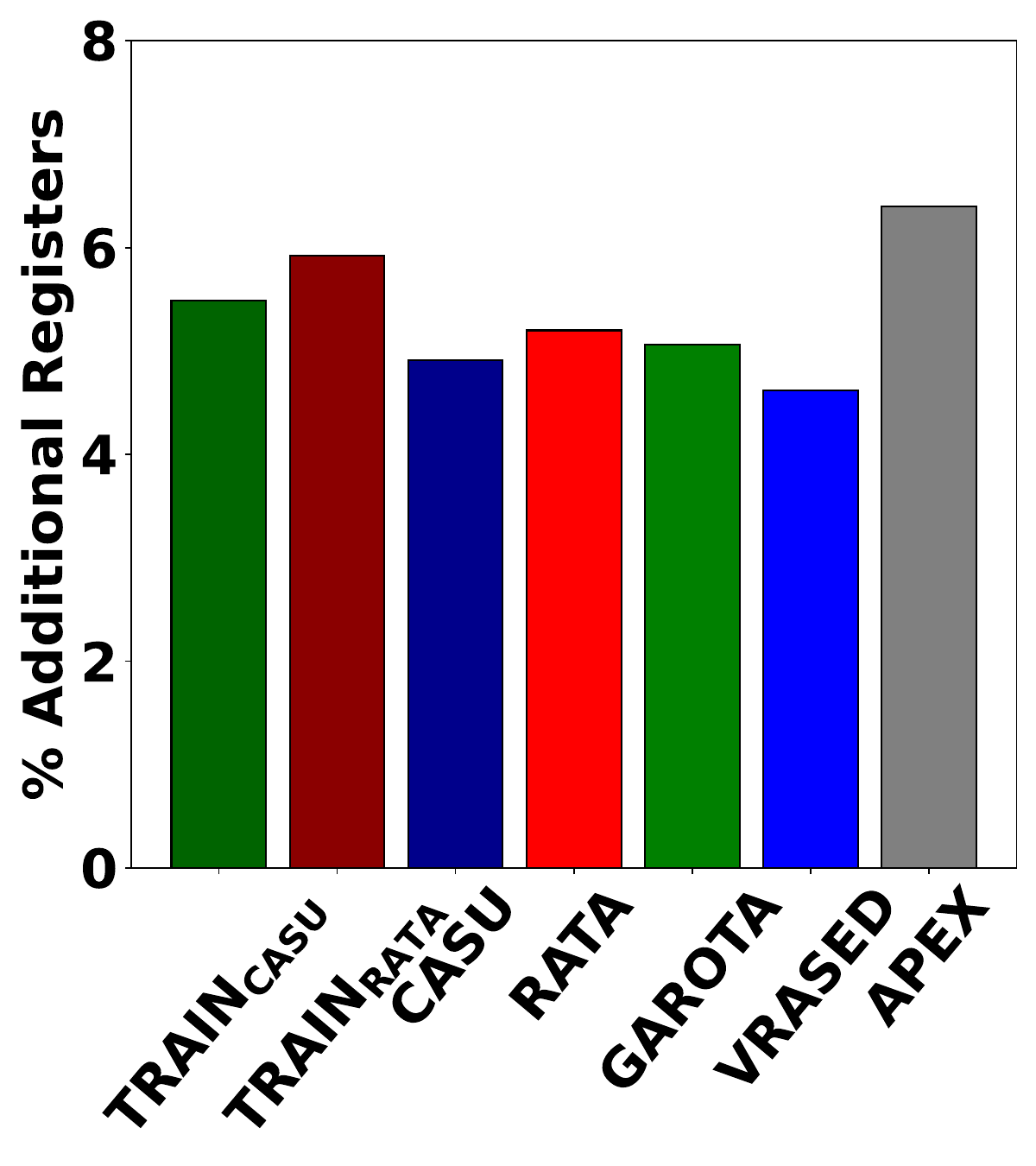}}
	\vspace{-1em}
	\caption{Hardware Overhead Comparison}
	\label{fig: hw_comparison} 
\end{figure}

\noindent {\bf Comparison with Other Hybrid \rot:}
We compare \system with other hybrid \rot constructions leveraging \ra:
VRASED \cite{vrased}, RATA \cite{rata}, CASU \cite{casu}, GAROTA \cite{garota}, and APEX \cite{nunes2020apex}. 
Note that RATA, CASU, APEX are implemented based on VRASED,
and all the above architectures are (in turn) based on openMSP430.
Results are shown in Figure \ref{fig: hw_comparison}.
APEX has a higher overhead than others due to additional hardware properties required
for generating proofs-of-execution.

\begin{table}  
  \centering\captionsetup{justification = centering}
  \footnotesize
  {
     \begin{tabular}{|Sc|Sc|Sc|} \hline
     \rowcolor{gray!20}
     \bf Architecture & \bf Look-Up Tables & \bf Registers  \\ \thickhline
     openMSP430 &  $1854$ & $692$ \\ \hline
     \casu & $1956$ & $726$ \\ \hline 
     \rowcolor{yellow!20}$\boldsymbol{\trapscasu}$ & {\bf $1967$ (+$11$)} & {\bf $740$ (+$14$)} \\ \hline
     \rata & $1928$ & $728$ \\ \hline 
     \rowcolor{yellow!20}$\boldsymbol{\trapsrata}$ & {\bf $1935$ (+$7$)} & {\bf $737$ (+$9$)} \\ \hline 
     \end{tabular}
  }
  \caption{\system Hardware Overhead} 
  \label{table: hardware overhead}
  \vspace{-3em}
\end{table}

\subsection{Run-time Overhead} \label{subsec:runtime-overhead}

Since \vrf is not a resource-constrained device, we focus on the overheads incurred on \prv.
Table \ref{table:run-time overhead} provides an overview of the run-time overhead for 
\system and a comparison with prominent prior \sa techniques: SEDA \cite{asokan2015seda}, 
SCRAPS \cite{petzi2022scraps}, DIAT \cite{abera2019diat}, and SANA \cite{ambrosin2016sana}.
\begin{table}
    
    \captionsetup{justification=centering}
    \scriptsize
    \resizebox{\columnwidth}{!}{
    \begin{tabular}{|c|c|c|}
        \hline
        \rowcolor{gray!20}
        & {\bf Request} & {\bf Report} \\
        \rowcolor{gray!20}
        & {\bf Verification} & {\bf Generation} \\
        \rowcolor{gray!20}
        \multirow{-3}{*}{\bf Architecture} & {\bf Time (ms)} & {\bf Time (ms)} \\
        \thickhline
        \makecell{\trapscasu  (@ $8$MHz)} &  $13.0$ & $29.5$ \\ \hline
     \makecell{\trapsrata  (@ $8$MHz)} & $12.9$ & $29.8$ \\ \hline 
     \makecell{SEDA Initiator \\ (SMART) (@ $8$MHz)} & N/A & \makecell{$56900$ $+$ \\ $256*g$} \\ \hline 
    \makecell{SEDA participating devices \\ (SMART) (@ $8$MHz)} & N/A & \makecell{$96$ $+$ \\ $256*(g-1)$}\\ \hline 
    \makecell{SEDA Initiator \\ (TRUSTLITE) (@ $24$MHz)} & N/A & $347.2 + 4.4*g$ \\ \hline 
    \makecell{SEDA participating devices \\ (TRUSTLITE) (@ $24$MHz)} & N/A & \makecell{$0.6$ $+$ \\ $4.4*(g-1)$} \\ \hline 
    \makecell{SCRAPS \\ (LPC55S69) (@ $150$MHz)} & N/A & $2109.1$ \\ \hline
    \makecell{SCRAPS \\ (ATmega1284P) (@ $20$MHz)} & N/A & $40147.4$ \\ \hline
    \makecell{DIAT  (@ $168$MHz)} & N/A & $835$ \\ \hline
    \makecell{SANA  (@ $48$MHz)} & $921.5$ & $3125.8$ \\ \hline
    \end{tabular}
    }
    \vspace{.05cm}
    \caption{\small Run-time Overhead Comparison 
    \\ ($g$: number of neighbors of a device)
    }
    \label{table:run-time overhead}
    \vspace{-4em}
\end{table}

Generating the attestation report (\Attrep) is quite fast for both \trapscasu\ and \trapsrata\ \prv types, since the overhead
is dominated by the computation of an HMAC over a minimal fixed-length region. 

In comparison, initiators in SEDA have 

to sign the entire  
aggregated report, resulting  in a  significantly longer timing overhead compared to \system. The report generation time of other \prv-s is also higher than \system as they must attest the whole program memory and verify neighbors' reports.

Moreover, report generation time in SEDA grows (almost) linearly, relying on the number of neighbors, denoted by $g$.

We also examine run-time overhead of SCRAPS, DIAT, and SANA.
These schemes perform relatively complex tasks as part of attestation and thus
incur high run-time overhead despite being implemented on more powerful devices. 

In summary, compared to DIAT, SCRAPS, and SANA, \system is lightweight in terms of run-time overhead.

\subsection{Energy Consumption}\label{subsec:energy-overhead}
\comment{
\begin{table}
        \footnotesize
        \centering
        \captionsetup{justification = centering}
        \begin{tabular}{|Sc|Sc|Sc|}
            \hline
            \rowcolor{gray!20}
            {\bf \small Device Type} & {\bf \small  Dynamic Power} & {\bf \small Device Static Power} \\
            \thickhline
            \makecell{\small \casu, \\ \rata}  & {\small $111 mW$} & {\small $72 mW$}\\
            \hline
            \makecell{\small \trapscasu, \\ \trapsrata}  & {\small $115 mW$} & {\small $72 mW$}\\
            \hline
        \end{tabular}  
        \vspace{.05cm}
        \caption{\system Energy Overhead}
        \label{table:power overhead}
\end{table}
Table \ref{table:power overhead} shows 
Estimated power consumption is measured by Xilinx Vivado.
}

Dynamic power consumption measurements from Xilinx Vivado show that \trapscasu{} and \trapsrata{} consume $115 mW$, of which $111 mW$ is consumed
by either \casu or \rata. This represents a $2$\% increase in total on-chip power.
Total time spent by \system (request verification and report generation) is $42.5$ms for \trapscasu and 
$42.7$ms for \trapsrata. Therefore, energy consumption per attestation instance is $\approx~0.00221$mWh 
for \trapscasu and \trapsrata, which is negligible.

\subsection{Scalability Eval via Network Simulation} \label{sec:network-simulation}

\begin{figure}
    \centering
    \captionsetup{justification=centering}
    \includegraphics[width=0.9\columnwidth]{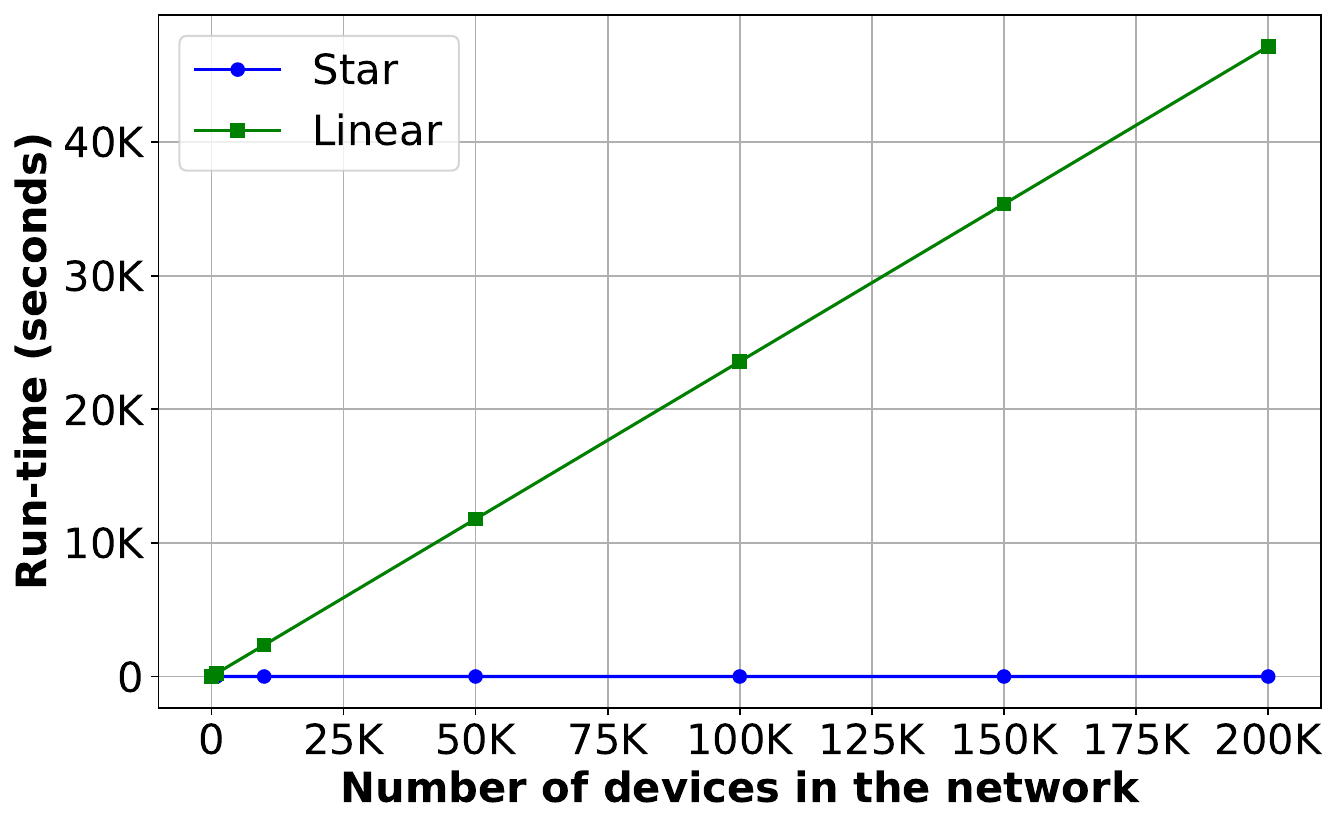}
    \caption{\system Simulation for Line/Star Topologies}
    \label{fig: star linear}
\end{figure}

\begin{figure}
    \centering
    \captionsetup{justification=centering}
    \includegraphics[width=0.9\columnwidth]{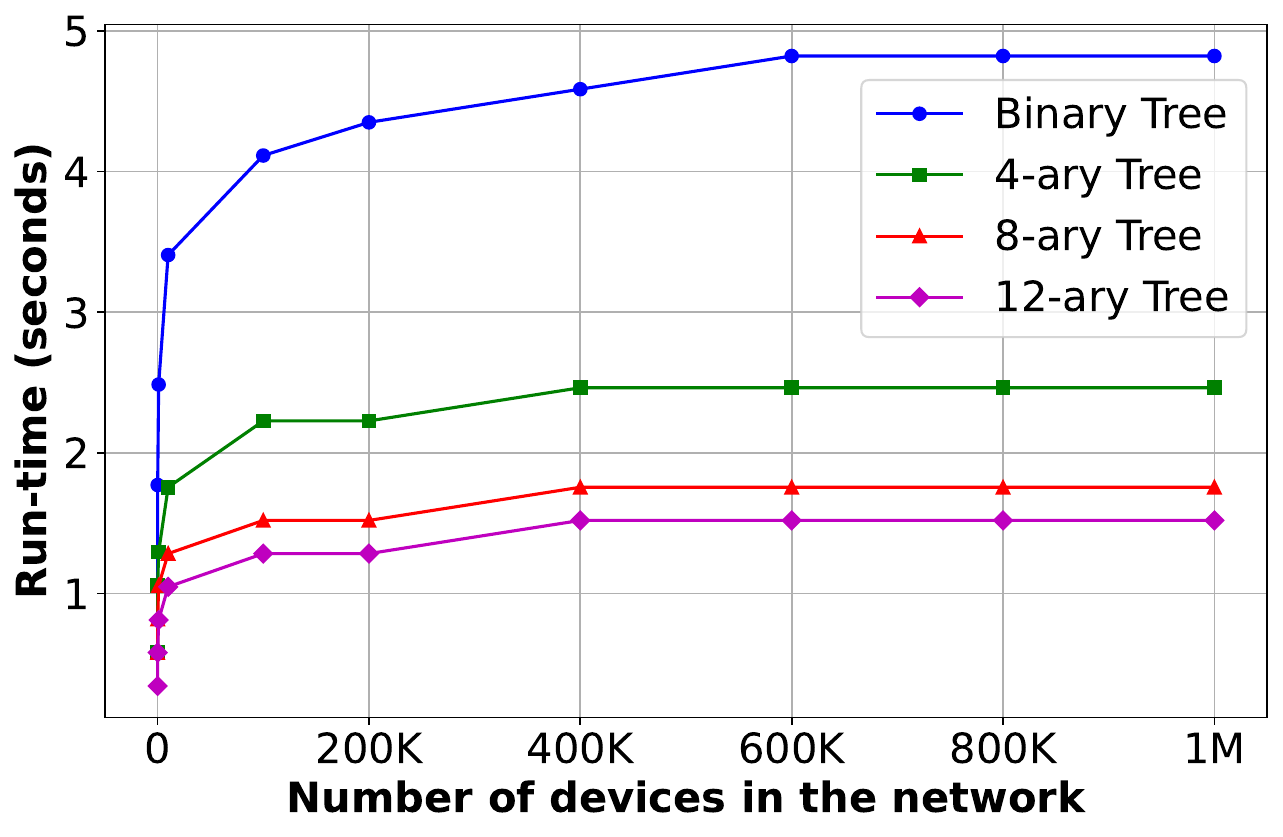}
    \caption{\system Simulation for Various Tree Topologies} 
    \label{fig: tree topologies}
\end{figure}

We conduct network simulations using the OMNeT++ \cite{omnetpp}. 
Since \trapsrtc and \trapsnortc protocols are similar, only the former is simulated.
Simulations are performed at the application layer. Cryptographic operations are simulated using delays 
that correspond to their actual execution times on \trapscasu and \trapsrata\ \prv-s.
We exclude \vrf's verification time from the simulations and set the communication rate between \prv-s to $250$Kbps.
This rate matches the standard data rate for ZigBee -- a common communication protocol for IoT devices.
Simulations are conducted with various spanning tree topologies: line, star, and several types of trees, with degrees
ranging from $2$ (binary) to $12$. We also vary the number of devices from $10$ to $1,000,000$.
Simulation results for \trapscasu and \trapsrata are almost identical, thus, only \trapscasu results are shown in Figures \ref{fig: star linear} and \ref{fig: tree topologies}.

As evident from Figure \ref{fig: star linear}, the run-time of \system is constant with the star topology and 
grows linearly with the linear topology.
This is because, in the former, \prv can start attestation almost immediately (as there is no forwarding to descendants), 
while each \prv waits until the farthest-away \prv is ready to perform attestation in the latter. 
The actual run-time for the star topology is $343$ms. For a network with a tree topology, 
\system run-time overhead is logarithmic in the number of \prv-s since the tree height governs it.
Simulation results show that \system is efficient in both small and large networks with various topologies. 
\section{Discussion \label{disc}}
\setlength\cellspacetoplimit{3pt}
\setlength\cellspacebottomlimit{3pt}

\begin{table*}[hbt!]
        \footnotesize
        \vspace{-2em}
        \centering\captionsetup{justification = centering}        
        {\begin{tabular}{|Sc|Sc|Sc|Sc|Sc|Sc|Sc|}
            \hline
            \rowcolor{gray!20}
            {\bf $\boldsymbol{\ra}$ Method} & {\bf Type} & {\bf  Passive/Active} & $\boldsymbol{\toctoura}$ & {\bf Network TCB} & {\bf Attestation Time} & {\bf Platform} \\            
            \thickhline
            RealSWATT \cite{surminski2021realswatt} & SW &
            Passive & \greencheck  & \redcross & $O(n)$ & ESP32 \\
            \hline
            PISTIS \cite{grisafi2022pistis} & SW &
            Passive & \redcross & \redcross & $O(n)$ & openMSP430 \\
            \hline
            SANCUS \cite{noorman2013sancus} & HW &
            Passive & \redcross & \redcross & $O(n)$ & openMSP430 \\
            \hline
            TrustVisor \cite{mccune2010trustvisor} & HW &
            Passive & \redcross & \redcross & $O(n)$ & AMD \\
            \hline
            VRASED \cite{vrased} & Hybrid &
            Passive & \redcross & \redcross & $O(n)$ & openMSP430 \\
            \hline
            IDA \cite{arkannezhadida} & Hybrid &
            Passive & \greencheck & \redcross & $O(n)$ & openMSP430 \\
            \hline
            RATA \cite{rata} & Hybrid &
            Passive & \greencheck & \redcross & $O(1)$ & openMSP430 \\
            \hline
            GAROTA \cite {garota} & Hybrid &
            Active & \redcross & \greencheck  & $O(n)$ & openMSP430 \\
            \hline
            CASU \cite {casu} & Hybrid &
            Active & \greencheck & \redcross & $O(1)$ & openMSP430 \\            
            \thickhline
            \rowcolor{yellow!20}
            {\bf TRAIN} & {\bf Hybrid} &
            {\bf Passive/Active} & \greencheckt & \greencheckt  & {\bf $O(1)$} & {\bf openMSP430} \\
            \hline
        \end{tabular}}
        \vspace{.1cm}
        \caption{Comparison with Other Individual Attestation Schemes ($n$: attested area size)}
        \label{table:comp_ind_att}
        \vspace{-1.5em}
\end{table*}
\begin{table*}[hbt!]
    \scriptsize
    \centering\captionsetup{justification = centering}
    {\centering}    \resizebox{\textwidth}{!}
    {\begin{tabular}{|Sc|Sc|Sc|Sc|Sc|}
        \hline
        \rowcolor{gray!20}
        {\bf $\boldsymbol{\sa}$ Method} & {\bf $\boldsymbol{\toctousa}$} & {\bf Simulator} & {\bf \makecell{Underlying \\ Platform}} & {\bf Remark} \\            
        \thickhline
        SEDA \cite{asokan2015seda} & \redcross &
        OMNeT++ & SMART/TrustLite & Provides pioneering scheme using secure hop-by-hop aggregation \\
        \hline
        SANA \cite{ambrosin2016sana} & \redcross &
        OMNeT++ & TyTan & Extends SEDA with aggregate signatures and sub-networks \\ 
        \hline
        LISA \cite{carpent2017lightweight} & \redcross &
        CORE & Unspecified & Introduces neighbor-based communication and quality metric \\ 
        \hline
        SeED \cite{ibrahim2017seed} & \redcross &
        OMNeT++ & SMART/TrustLite & Extends SEDA with self-initiated \ra\\ 
        \thickhline
        DARPA \cite{ibrahim2016darpa} & \redcross &
        OMNeT++ & SMART & Exchanges heartbeat messages to detect physically compromised devices \\
        \hline
        SCAPI \cite{kohnhauser2017scapi} & \redcross &
        OMNeT++ & ARM Cortex-M4 & Extends DARPA with regular session key generation and distribution on \prv-s \\
        \hline
        SAP \cite{nunes2019towards} & \redcross &
        OMNeT++ & TrustLite & Constructs formal model with security notions for \sa \\ 
        \hline
        SALAD \cite {kohnhauser2018salad} & \redcross &
        OMNeT++ & ARM Cortex-M4 & Offers lightweight message aggregation in dynamic topology \\ 
        \hline
        SCRAPS \cite {petzi2022scraps} & \redcross &
        Python-based & ARM Cortex-M33 & Constructs Pub/Sub protocol using blockchain-hosted smart contracts\\
        \hline
        ESDRA \cite {kuang2019esdra} & \redcross &
        OMNeT++ & Unspecified & Presents many-to-one \sa scheme to eliminate fixed \vrf \\
        \hline
        DIAT \cite {abera2019diat} & \redcross &
        OMNeT++ & PX4 & Introduces control-flow attestation for autonomous collaborative systems \\
        \hline
        \rowcolor{yellow!20}
        {\bf TRAIN} & \greencheckt &
        {\bf OMNeT++} & {\bf CASU/RATA} & Minimizes \toctou window, \ra overhead, and isolates \ra functionality\\
        \hline
    \end{tabular}}      
    \vspace{.1cm}
    \caption{Comparison with Other Network Attestation Schemes}
    \vspace{-1em}
    \label{table:comp_swa_att}
\end{table*}

\subsection{\system Compatibility}
The rationale behind our choice of \prv\ \ra platforms (i.e., \casu and \rata) is due to their minimal
\ra\ overhead (HMAC over minimal fixed size input), \toctoura\ mitigation, and extensibility,
which facilitates the construction of TimerTCB and NetTCB with a few hardware modifications.
However, \system is also compatible with other \ra platforms to minimize the \toctousa window,
while losing the benefits of \casu and \rata.
Some examples of compatible devices are:
\begin{compactitem}
    \item Devices with custom hardware \rot, e.g., Sancus~\cite{noorman2013sancus} or TrustVisor~\cite{mccune2010trustvisor}.
    \item Devices with off-the-shelf TEE, such as TrustZone-A or TrustZone-M \cite{trustzone}.
    \item Devices with hybrid (HW/SW) \rot, such as SMART~\cite{smart}, VRASED~\cite{vrased}, TyTAN~\cite{tytan}, TrustLite~\cite{trustlite}.
    \item Devices without any hardware \rot. In this case, the device OS must be trusted.
\end{compactitem}

\subsection{\rata vs. \casu} \label{subsec:rata-casu-comparison}
Given that \rata operates as a passive \rot and \casu functions as an active \rot, 
it is natural to question the necessity of \rata and why \casu is not utilized exclusively.
The justification for employing \rata over \casu stems from three primary reasons:
(1) Memory Constraints: In \casu, only half of the program memory (PMEM) can store authorized software, while the other half is reserved for the secure update process. This significant (50\%) PMEM reservation can be prohibitive for low-end devices with limited memory. 
(2) Access to Non-Volatile Memory: \casu prevents normal software from modifying PMEM. However, some software may require access to non-volatile memory (e.g., flash) for benign purposes, such as storing text or image files. \rata allows such access and is preferred in these circumstances.
(3) Hardware Overheads: \rata has slightly lower hardware overheads compared to \casu. 

\subsection{\toctousa Minimization in \trapsnortc }\label{dis:precise_sync}
Even though \trapsnortc cannot achieve perfect synchronization without RTCs, it significantly 
reduces the \toctousa window compared to na\"ive approaches where the window scales with both 
spanning tree depth and network congestion. Recall that Section~\ref{subsec:security-analysis} illustrates the reduction in \toctousa window through a concrete example. By computing attestation timing based on the 
network topology, \trapsnortc effectively eliminates the spanning tree traversal component 
of the \toctousa window, leaving only network delay as a factor influencing the
imperfection of the synchronization. 

\section{Related Work \label{related}}
\noindent~{\bf Individual Device Attestation (\ra)}
is an extensively studied topic and numerous schemes have been proposed in the literature. These 
techniques generally fall into three categories: software-based, hardware-based, and hybrid.
Given the lack of rich hardware features on embedded platforms, lightweight Software-based \ra 
\cite{li2011viper,seshadri2006scuba,seshadri2004swatt,surminski2021realswatt} is only viable 
for legacy devices with no secure hardware features. 
It uses request-to-response time (between \vrf and \prv) to establish confidence 
in the integrity of the attestation report.
Nonetheless, network limitations (e.g. intermittent connection, network congestion) on \prv introduce noise to the request-to-response 
time, making software-based \ra impractical.

In contrast, hardware-based \ra techniques 
\cite{mccune2010trustvisor,noorman2013sancus,strackx2010efficient,
ling2021secure,chen2019opera,chen2022mage} either (1) embed \prv attestation functionality entirely 
within dedicated hardware, 
or (2) require substantial changes to the underlying hardware to support isolated execution of trusted 
software, e.g., SGX \cite{sgx} or TrustZone \cite{trustzone}.
However, such hardware features are often too complex and costly for low-end devices constrained by 
size, energy, and cost.

Given the limitations of both hardware- and software-based approaches in low-end embedded platforms, 
software/hardware co-design (hybrid) \cite{vrased,arkannezhadida,smart,tytan,nunes2020apex,trustlite} 
has recently emerged as a promising solution. It aims to provide equivalent security guarantees to hardware-based 
\ra while minimizing modifications to the underlying hardware.
The security features employed can be simplified to utilize a ROM or a memory protection unit (MPU).
Current hybrid \ra techniques implement the integrity-ensuring function (e.g., MAC) in software.
They use trusted hardware to control the execution of this software, preventing any violations that may compromise security, such as key leakage, or preemption of unprivileged software.

RealSWATT \cite{surminski2021realswatt} introduces a software-based approach designed for continuous attestation of real-time and multi-core systems, effectively solving the TOCTOU problem.
PISTIS \cite{grisafi2022pistis} is also a software trusted computing architecture enabling memory isolation, remote attestation, and secure update.
SANCUS \cite{noorman2013sancus} and TrustVisor \cite{mccune2010trustvisor} are hardware-based solutions offering attestation service with software module isolation. 
VRASED \cite{vrased} presents a formally verified hybrid RA architecture.
It implements the attestation function in software while employing small trusted hardware to enforce the attestation correctness and access control over the \ra secret key.
IDA \cite{arkannezhadida} proposes a novel hybrid attestation method that enables interrupts even during attestation, enhancing overall system security and flexibility.
Moreover, IDA monitors program memory between attestation requests to prevent TOCTOU attacks.
As previously mentioned in Section \ref{sec:bg}, \rata, \casu, and \garota are hybrid \ra architectures.
The first two provide constant-time computation for attestation requests (heartbeat requests in \casu) regardless of the size of the attested regions.
Meanwhile, the last provides a trusted timer and network that can be preemptively executed by authorized software.
Table \ref{table:comp_ind_att} compares various software, hardware, and hybrid \ra methods.

\noindent~{\bf Network Attestation (\sa)}
enables scalable attestation for large groups of interconnected devices. Few prior work \cite{asokan2015seda,ambrosin2016sana,carpent2017lightweight,ibrahim2017seed,ibrahim2016darpa,kohnhauser2017scapi,nunes2019towards,kohnhauser2018salad,petzi2022scraps,kuang2019esdra,abera2019diat} refers to this process as Swarm Attestation; we employ the term Network Attestation to denote the same concept. Table \ref{table:comp_swa_att} shows a comparison with other \sa schemes.

The first scheme, SEDA \cite{asokan2015seda}, employs secure hop-by-hop aggregation of \ra reports. 
Initially, \vrf broadcasts an attestation request to \prv-s. Each \prv attests its children nodes and forwards aggregated \ra reports to its parent. Finally, \vrf verifies only the last \ra reports to assess the status of all \prv-s.
SANA \cite{ambrosin2016sana} extends SEDA with a novel aggregate signature scheme, ensuring low verification overhead with minimal trust anchor.
It partitions \prv-s into subnetworks and aggregates \ra results across the entire network, facilitating public verification by multiple \vrf-s.
LISA \cite{carpent2017lightweight} introduces neighbor-based communication to propagate \ra reports. \prv-s verify \ra reports of other \prv-s before forwarding them to prevent replay attacks, and a quality metric for \sa techniques captures the information from each \prv.
SeED \cite{ibrahim2017seed} enhances the efficiency of SEDA and resilience against DoS attacks by enabling \prv-s to self-initiate \ra.
DARPA \cite{ibrahim2016darpa} detects physically compromised devices by exchanging heartbeat messages among \prv-s to identify compromised or absent devices.
SCAPI \cite{kohnhauser2017scapi} improves DARPA;  it introduces a leader that periodically generates and distributes secret session keys among \prv-s. To receive a new session key, \prv must be authenticated with the previous key.
SAP \cite{nunes2019towards} constructs a formal model encompassing desirable efficiency, soundness, and security notions for \sa. It systematically designs a synchronous attestation protocol compliant with security goals defined by the formal model.
SALAD \cite{kohnhauser2018salad} provides lightweight message aggregation for dynamic networks with intermittent connectivity, distributing \ra proofs among all devices.

SCRAPS \cite{petzi2022scraps} proposes a Pub/Sub network \sa protocol. It involves a proxy verifying \prv’s \ra reports on behalf of actual \vrf.
This proxy is implemented using smart contracts, i.e., untrusted entities hosted on a blockchain.
Once the proxy attests a \prv, \vrf-s can retrieve the \ra evidence from the proxy without trusting the proxy, enabling many-to-many attestation.
This enables many-to-many attestation by allowing \vrf-s to fetch \ra reports from the proxy.
ESDRA \cite{kuang2019esdra} designs a first many-to-one \sa scheme to eliminate fixed \vrf and reduce a single point of failure \vrf risks.
Moreover, the distributed attestation facilitates offering feedback on certain compromised \prv-s, thus suitable for half-dynamic networks.
DIAT \cite{abera2019diat} presents a control-flow attestation scheme for autonomous collaborative systems.
It combines data integrity attestation, modular attestation, and representation of execution paths, enabling efficient run-time attestation in a setting where embedded systems must act as both, \prv and \vrf.

\section{Conclusions \label{conc}}

This paper constructs a \toctou-resilient \sa protocol (\system) 
for networks of low-end IoT devices. It facilitates simultaneous attestation across the network while 
minimizing runtime/energy overhead by computing HMAC over minimal fixed-size input. 
Two variants of the protocol, based on the availability of real-time clocks, are present.

An open-source prototype implemented on TI MSP430 demonstrates the practicality of \system on commodity hardware.

\bibliographystyle{ACM-Reference-Format}
\bibliography{reference}


\begin{thebibliography}{58}


\ifx \showCODEN    \undefined \def \showCODEN     #1{\unskip}     \fi
\ifx \showDOI      \undefined \def \showDOI       #1{#1}\fi
\ifx \showISBNx    \undefined \def \showISBNx     #1{\unskip}     \fi
\ifx \showISBNxiii \undefined \def \showISBNxiii  #1{\unskip}     \fi
\ifx \showISSN     \undefined \def \showISSN      #1{\unskip}     \fi
\ifx \showLCCN     \undefined \def \showLCCN      #1{\unskip}     \fi
\ifx \shownote     \undefined \def \shownote      #1{#1}          \fi
\ifx \showarticletitle \undefined \def \showarticletitle #1{#1}   \fi
\ifx \showURL      \undefined \def \showURL       {\relax}        \fi
\providecommand\bibfield[2]{#2}
\providecommand\bibinfo[2]{#2}
\providecommand\natexlab[1]{#1}
\providecommand\showeprint[2][]{arXiv:#2}

\bibitem[Abera et~al\mbox{.}(2019)]%
        {abera2019diat}
\bibfield{author}{\bibinfo{person}{Tigist Abera}, \bibinfo{person}{Raad Bahmani}, \bibinfo{person}{Ferdinand Brasser}, \bibinfo{person}{Ahmad Ibrahim}, \bibinfo{person}{Ahmad-Reza Sadeghi}, {and} \bibinfo{person}{Matthias Schunter}.} \bibinfo{year}{2019}\natexlab{}.
\newblock \showarticletitle{DIAT: Data Integrity Attestation for Resilient Collaboration of Autonomous Systems.}. In \bibinfo{booktitle}{\emph{NDSS}}.
\newblock


\bibitem[Aldoseri et~al\mbox{.}(2023)]%
        {aldoseri2023symbolic}
\bibfield{author}{\bibinfo{person}{Abdulla Aldoseri}, \bibinfo{person}{Tom Chothia}, \bibinfo{person}{Jose Moreira}, {and} \bibinfo{person}{David Oswald}.} \bibinfo{year}{2023}\natexlab{}.
\newblock \showarticletitle{Symbolic modelling of remote attestation protocols for device and app integrity on Android}. In \bibinfo{booktitle}{\emph{Proceedings of the 2023 ACM Asia Conference on Computer and Communications Security}}.
\newblock


\bibitem[Aliaj et~al\mbox{.}(2022)]%
        {garota}
\bibfield{author}{\bibinfo{person}{Esmerald Aliaj}, \bibinfo{person}{Ivan De~Oliveira Nunes}, {and} \bibinfo{person}{Gene Tsudik}.} \bibinfo{year}{2022}\natexlab{}.
\newblock \showarticletitle{$\{$GAROTA$\}$: generalized active $\{$Root-Of-Trust$\}$ architecture (for tiny embedded devices)}. In \bibinfo{booktitle}{\emph{31st USENIX Security Symposium (USENIX Security 22)}}.
\newblock


\bibitem[Ambrosin et~al\mbox{.}(2016)]%
        {ambrosin2016sana}
\bibfield{author}{\bibinfo{person}{Moreno Ambrosin}, \bibinfo{person}{Mauro Conti}, \bibinfo{person}{Ahmad Ibrahim}, \bibinfo{person}{Gregory Neven}, \bibinfo{person}{Ahmad-Reza Sadeghi}, {and} \bibinfo{person}{Matthias Schunter}.} \bibinfo{year}{2016}\natexlab{}.
\newblock \showarticletitle{SANA: Secure and scalable aggregate network attestation}. In \bibinfo{booktitle}{\emph{Proceedings of the 2016 ACM SIGSAC conference on computer and communications security}}.
\newblock


\bibitem[Arkannezhad et~al\mbox{.}(2024)]%
        {arkannezhadida}
\bibfield{author}{\bibinfo{person}{Fatemeh Arkannezhad}, \bibinfo{person}{Justin Feng}, {and} \bibinfo{person}{Nader Sehatbakhsh}.} \bibinfo{year}{2024}\natexlab{}.
\newblock \showarticletitle{IDA: Hybrid Attestation with Support for Interrupts and TOCTOU}. In \bibinfo{booktitle}{\emph{31th Annual Network and Distributed System Security Symposium, NDSS 2024}}.
\newblock


\bibitem[{Arm Ltd.}(2018)]%
        {trustzone}
\bibfield{author}{\bibinfo{person}{{Arm Ltd.}}} \bibinfo{year}{2018}\natexlab{}.
\newblock \bibinfo{title}{Arm {TrustZone}}.
\newblock \bibinfo{howpublished}{\url{https://www.arm.com/products/security-on-arm/trustzone/}}.
\newblock


\bibitem[Asokan et~al\mbox{.}(2015)]%
        {asokan2015seda}
\bibfield{author}{\bibinfo{person}{Nadarajah Asokan}, \bibinfo{person}{Ferdinand Brasser}, \bibinfo{person}{Ahmad Ibrahim}, \bibinfo{person}{Ahmad-Reza Sadeghi}, \bibinfo{person}{Matthias Schunter}, \bibinfo{person}{Gene Tsudik}, {and} \bibinfo{person}{Christian Wachsmann}.} \bibinfo{year}{2015}\natexlab{}.
\newblock \showarticletitle{Seda: Scalable embedded device attestation}. In \bibinfo{booktitle}{\emph{Proceedings of the 22nd ACM SIGSAC Conference on Computer and Communications Security}}.
\newblock


\bibitem[Brasser et~al\mbox{.}(2015)]%
        {tytan}
\bibfield{author}{\bibinfo{person}{Ferdinand Brasser}, \bibinfo{person}{Brahim~El Mahjoub}, \bibinfo{person}{Ahmad{-}Reza Sadeghi}, \bibinfo{person}{Christian Wachsmann}, {and} \bibinfo{person}{Patrick Koeberl}.} \bibinfo{year}{2015}\natexlab{}.
\newblock \showarticletitle{TyTAN: tiny trust anchor for tiny devices}. In \bibinfo{booktitle}{\emph{Proceedings of the 52nd Annual Design Automation Conference, San Francisco, CA, USA, June 7-11, 2015}}.
\newblock


\bibitem[Brasser et~al\mbox{.}(2016)]%
        {Brasser-DAC16}
\bibfield{author}{\bibinfo{person}{Ferdinand Brasser}, \bibinfo{person}{Kasper~Bonne Rasmussen}, \bibinfo{person}{Ahmad{-}Reza Sadeghi}, {and} \bibinfo{person}{Gene Tsudik}.} \bibinfo{year}{2016}\natexlab{}.
\newblock \showarticletitle{Remote attestation for low-end embedded devices: the prover's perspective}. In \bibinfo{booktitle}{\emph{Proceedings of the 53rd Annual Design Automation Conference, {DAC} 2016, Austin, TX, USA, June 5-9, 2016}}.
\newblock


\bibitem[Carpent et~al\mbox{.}(2017)]%
        {carpent2017lightweight}
\bibfield{author}{\bibinfo{person}{Xavier Carpent}, \bibinfo{person}{Karim ElDefrawy}, \bibinfo{person}{Norrathep Rattanavipanon}, {and} \bibinfo{person}{Gene Tsudik}.} \bibinfo{year}{2017}\natexlab{}.
\newblock \showarticletitle{Lightweight swarm attestation: A tale of two lisa-s}. In \bibinfo{booktitle}{\emph{Proceedings of the 2017 ACM on Asia Conference on Computer and Communications Security}}.
\newblock


\bibitem[Chen and Zhang(2022)]%
        {chen2022mage}
\bibfield{author}{\bibinfo{person}{Guoxing Chen} {and} \bibinfo{person}{Yinqian Zhang}.} \bibinfo{year}{2022}\natexlab{}.
\newblock \showarticletitle{$\{$MAGE$\}$: Mutual Attestation for a Group of Enclaves without Trusted Third Parties}. In \bibinfo{booktitle}{\emph{31st USENIX Security Symposium (USENIX Security 22)}}.
\newblock


\bibitem[Chen et~al\mbox{.}(2019)]%
        {chen2019opera}
\bibfield{author}{\bibinfo{person}{Guoxing Chen}, \bibinfo{person}{Yinqian Zhang}, {and} \bibinfo{person}{Ten-Hwang Lai}.} \bibinfo{year}{2019}\natexlab{}.
\newblock \showarticletitle{Opera: Open remote attestation for intel's secure enclaves}. In \bibinfo{booktitle}{\emph{Proceedings of the 2019 ACM SIGSAC Conference on Computer and Communications Security}}.
\newblock


\bibitem[Cimatti et~al\mbox{.}(2002)]%
        {cimatti2002nusmv}
\bibfield{author}{\bibinfo{person}{Alessandro Cimatti}, \bibinfo{person}{Edmund Clarke}, \bibinfo{person}{Enrico Giunchiglia}, \bibinfo{person}{Fausto Giunchiglia}, \bibinfo{person}{Marco Pistore}, \bibinfo{person}{Marco Roveri}, \bibinfo{person}{Roberto Sebastiani}, {and} \bibinfo{person}{Armando Tacchella}.} \bibinfo{year}{2002}\natexlab{}.
\newblock \showarticletitle{Nusmv 2: An opensource tool for symbolic model checking}. In \bibinfo{booktitle}{\emph{Computer Aided Verification: 14th International Conference, CAV 2002 Copenhagen, Denmark, July 27--31, 2002 Proceedings 14}}. Springer, \bibinfo{pages}{359--364}.
\newblock


\bibitem[De~Oliveira~Nunes et~al\mbox{.}(2022)]%
        {casu}
\bibfield{author}{\bibinfo{person}{Ivan De~Oliveira~Nunes}, \bibinfo{person}{Sashidhar Jakkamsetti}, \bibinfo{person}{Youngil Kim}, {and} \bibinfo{person}{Gene Tsudik}.} \bibinfo{year}{2022}\natexlab{}.
\newblock \showarticletitle{Casu: Compromise avoidance via secure update for low-end embedded systems}. In \bibinfo{booktitle}{\emph{Proceedings of the 41st IEEE/ACM International Conference on Computer-Aided Design}}.
\newblock


\bibitem[De~Oliveira~Nunes et~al\mbox{.}(2021)]%
        {rata}
\bibfield{author}{\bibinfo{person}{Ivan De~Oliveira~Nunes}, \bibinfo{person}{Sashidhar Jakkamsetti}, \bibinfo{person}{Norrathep Rattanavipanon}, {and} \bibinfo{person}{Gene Tsudik}.} \bibinfo{year}{2021}\natexlab{}.
\newblock \showarticletitle{On the TOCTOU problem in remote attestation}. In \bibinfo{booktitle}{\emph{Proceedings of the 2021 ACM SIGSAC Conference on Computer and Communications Security}}.
\newblock


\bibitem[Dolev and Yao(1983)]%
        {dolevYao}
\bibfield{author}{\bibinfo{person}{D. Dolev} {and} \bibinfo{person}{A. Yao}.} \bibinfo{year}{1983}\natexlab{}.
\newblock \showarticletitle{On the security of public key protocols}.
\newblock \bibinfo{journal}{\emph{IEEE Transactions on Information Theory}} (\bibinfo{year}{1983}).
\newblock


\bibitem[Eldefrawy et~al\mbox{.}(2012)]%
        {smart}
\bibfield{author}{\bibinfo{person}{Karim Eldefrawy}, \bibinfo{person}{Gene Tsudik}, \bibinfo{person}{Aur{\'e}lien Francillon}, {and} \bibinfo{person}{Daniele Perito}.} \bibinfo{year}{2012}\natexlab{}.
\newblock \showarticletitle{{SMART}: Secure and Minimal Architecture for (Establishing Dynamic) Root of Trust}. In \bibinfo{booktitle}{\emph{NDSS}}.
\newblock


\bibitem[Evtyushkin et~al\mbox{.}(2014)]%
        {evtyushkin2014iso}
\bibfield{author}{\bibinfo{person}{Dmitry Evtyushkin}, \bibinfo{person}{Jesse Elwell}, \bibinfo{person}{Meltem Ozsoy}, \bibinfo{person}{Dmitry Ponomarev}, \bibinfo{person}{Nael~Abu Ghazaleh}, {and} \bibinfo{person}{Ryan Riley}.} \bibinfo{year}{2014}\natexlab{}.
\newblock \showarticletitle{Iso-x: A flexible architecture for hardware-managed isolated execution}. In \bibinfo{booktitle}{\emph{2014 47th Annual IEEE/ACM International Symposium on Microarchitecture}}.
\newblock


\bibitem[Feng et~al\mbox{.}(2021)]%
        {feng2021scalable}
\bibfield{author}{\bibinfo{person}{Erhu Feng}, \bibinfo{person}{Xu Lu}, \bibinfo{person}{Dong Du}, \bibinfo{person}{Bicheng Yang}, \bibinfo{person}{Xueqiang Jiang}, \bibinfo{person}{Yubin Xia}, \bibinfo{person}{Binyu Zang}, {and} \bibinfo{person}{Haibo Chen}.} \bibinfo{year}{2021}\natexlab{}.
\newblock \showarticletitle{Scalable memory protection in the $\{$PENGLAI$\}$ enclave}. In \bibinfo{booktitle}{\emph{15th $\{$USENIX$\}$ Symposium on Operating Systems Design and Implementation ($\{$OSDI$\}$ 21)}}.
\newblock


\bibitem[Ghaeini et~al\mbox{.}(2019)]%
        {ghaeini2019patt}
\bibfield{author}{\bibinfo{person}{Hamid~Reza Ghaeini}, \bibinfo{person}{Matthew Chan}, \bibinfo{person}{Raad Bahmani}, \bibinfo{person}{Ferdinand Brasser}, \bibinfo{person}{Luis Garcia}, \bibinfo{person}{Jianying Zhou}, \bibinfo{person}{Ahmad-Reza Sadeghi}, \bibinfo{person}{Nils~Ole Tippenhauer}, {and} \bibinfo{person}{Saman Zonouz}.} \bibinfo{year}{2019}\natexlab{}.
\newblock \showarticletitle{$\{$PAtt$\}$: Physics-based Attestation of Control Systems}. In \bibinfo{booktitle}{\emph{22nd International Symposium on Research in Attacks, Intrusions and Defenses (RAID 2019)}}.
\newblock


\bibitem[Grisafi et~al\mbox{.}(2022)]%
        {grisafi2022pistis}
\bibfield{author}{\bibinfo{person}{Michele Grisafi}, \bibinfo{person}{Mahmoud Ammar}, \bibinfo{person}{Marco Roveri}, {and} \bibinfo{person}{Bruno Crispo}.} \bibinfo{year}{2022}\natexlab{}.
\newblock \showarticletitle{$\{$PISTIS$\}$: Trusted Computing Architecture for Low-end Embedded Systems}. In \bibinfo{booktitle}{\emph{31st USENIX Security Symposium (USENIX Security 22)}}.
\newblock


\bibitem[Ibrahim et~al\mbox{.}(2016)]%
        {ibrahim2016darpa}
\bibfield{author}{\bibinfo{person}{Ahmad Ibrahim}, \bibinfo{person}{Ahmad-Reza Sadeghi}, \bibinfo{person}{Gene Tsudik}, {and} \bibinfo{person}{Shaza Zeitouni}.} \bibinfo{year}{2016}\natexlab{}.
\newblock \showarticletitle{Darpa: Device attestation resilient to physical attacks}. In \bibinfo{booktitle}{\emph{Proceedings of the 9th ACM Conference on Security \& Privacy in Wireless and Mobile Networks}}.
\newblock


\bibitem[Ibrahim et~al\mbox{.}(2017)]%
        {ibrahim2017seed}
\bibfield{author}{\bibinfo{person}{Ahmad Ibrahim}, \bibinfo{person}{Ahmad-Reza Sadeghi}, {and} \bibinfo{person}{Shaza Zeitouni}.} \bibinfo{year}{2017}\natexlab{}.
\newblock \showarticletitle{SeED: secure non-interactive attestation for embedded devices}. In \bibinfo{booktitle}{\emph{Proceedings of the 10th ACM conference on security and privacy in wireless and mobile networks}}.
\newblock


\bibitem[Intel({[n.\,d.]})]%
        {sgx}
\bibfield{author}{\bibinfo{person}{Intel}.} \bibinfo{year}{[n.\,d.]}\natexlab{}.
\newblock \bibinfo{title}{{Software Guard Extensions} ({Intel} {SGX})}.
\newblock \bibinfo{howpublished}{\url{https://software.intel.com/en-us/sgx/}}.
\newblock


\bibitem[Irfan et~al\mbox{.}(2016)]%
        {irfan2016verilog2smv}
\bibfield{author}{\bibinfo{person}{Ahmed Irfan}, \bibinfo{person}{Alessandro Cimatti}, \bibinfo{person}{Alberto Griggio}, \bibinfo{person}{Marco Roveri}, {and} \bibinfo{person}{Roberto Sebastiani}.} \bibinfo{year}{2016}\natexlab{}.
\newblock \showarticletitle{Verilog2SMV: A tool for word-level verification}. In \bibinfo{booktitle}{\emph{2016 Design, Automation \& Test in Europe Conference \& Exhibition (DATE)}}. IEEE, \bibinfo{pages}{1156--1159}.
\newblock


\bibitem[J.-K.~Zinzindohoué and Beurdouche(2017)]%
        {hacl}
\bibfield{author}{\bibinfo{person}{J.~Protzenko J.-K.~Zinzindohoué, K.~Bhargavan} {and} \bibinfo{person}{B. Beurdouche}.} \bibinfo{year}{2017}\natexlab{}.
\newblock \showarticletitle{“Hacl*: A verified modern cryptographic library}, In \bibinfo{booktitle}{“Hacl*: A verified modern cryptographic library}.
\newblock \bibinfo{journal}{\emph{CCS}}.
\newblock


\bibitem[Jakkamsetti et~al\mbox{.}(2023)]%
        {jakkamsetti2023caveat}
\bibfield{author}{\bibinfo{person}{Sashidhar Jakkamsetti}, \bibinfo{person}{Youngil Kim}, {and} \bibinfo{person}{Gene Tsudik}.} \bibinfo{year}{2023}\natexlab{}.
\newblock \showarticletitle{Caveat (IoT) Emptor: Towards Transparency of IoT Device Presence}. In \bibinfo{booktitle}{\emph{Proceedings of the 2023 ACM SIGSAC Conference on Computer and Communications Security}}.
\newblock


\bibitem[Koeberl et~al\mbox{.}(2014)]%
        {trustlite}
\bibfield{author}{\bibinfo{person}{Patrick Koeberl}, \bibinfo{person}{Steffen Schulz}, \bibinfo{person}{Ahmad-Reza Sadeghi}, {and} \bibinfo{person}{Vijay Varadharajan}.} \bibinfo{year}{2014}\natexlab{}.
\newblock \showarticletitle{{TrustLite}: A security architecture for tiny embedded devices}. In \bibinfo{booktitle}{\emph{EuroSys}}.
\newblock


\bibitem[Kohnh{\"a}user et~al\mbox{.}(2017)]%
        {kohnhauser2017scapi}
\bibfield{author}{\bibinfo{person}{Florian Kohnh{\"a}user}, \bibinfo{person}{Niklas B{\"u}scher}, \bibinfo{person}{Sebastian Gabmeyer}, {and} \bibinfo{person}{Stefan Katzenbeisser}.} \bibinfo{year}{2017}\natexlab{}.
\newblock \showarticletitle{Scapi: a scalable attestation protocol to detect software and physical attacks}. In \bibinfo{booktitle}{\emph{Proceedings of the 10th ACM conference on security and privacy in wireless and mobile networks}}.
\newblock


\bibitem[Kohnh{\"a}user et~al\mbox{.}(2018)]%
        {kohnhauser2018salad}
\bibfield{author}{\bibinfo{person}{Florian Kohnh{\"a}user}, \bibinfo{person}{Niklas B{\"u}scher}, {and} \bibinfo{person}{Stefan Katzenbeisser}.} \bibinfo{year}{2018}\natexlab{}.
\newblock \showarticletitle{Salad: Secure and lightweight attestation of highly dynamic and disruptive networks}. In \bibinfo{booktitle}{\emph{Proceedings of the 2018 on Asia Conference on Computer and Communications Security}}.
\newblock


\bibitem[Kuang et~al\mbox{.}(2019)]%
        {kuang2019esdra}
\bibfield{author}{\bibinfo{person}{Boyu Kuang}, \bibinfo{person}{Anmin Fu}, \bibinfo{person}{Shui Yu}, \bibinfo{person}{Guomin Yang}, \bibinfo{person}{Mang Su}, {and} \bibinfo{person}{Yuqing Zhang}.} \bibinfo{year}{2019}\natexlab{}.
\newblock \showarticletitle{ESDRA: An efficient and secure distributed remote attestation scheme for IoT swarms}.
\newblock \bibinfo{journal}{\emph{IEEE Internet of Things Journal}} (\bibinfo{year}{2019}).
\newblock


\bibitem[Lamport(1981)]%
        {lamport1981}
\bibfield{author}{\bibinfo{person}{Leslie Lamport}.} \bibinfo{year}{1981}\natexlab{}.
\newblock \showarticletitle{Password Authentication with Insecure Communication}. In \bibinfo{booktitle}{\emph{Communications of the ACM 24.11}}.
\newblock


\bibitem[Li et~al\mbox{.}(2011)]%
        {li2011viper}
\bibfield{author}{\bibinfo{person}{Yanlin Li}, \bibinfo{person}{Jonathan~M McCune}, {and} \bibinfo{person}{Adrian Perrig}.} \bibinfo{year}{2011}\natexlab{}.
\newblock \showarticletitle{VIPER: Verifying the integrity of peripherals' firmware}. In \bibinfo{booktitle}{\emph{Proceedings of the 18th ACM conference on Computer and communications security}}.
\newblock


\bibitem[Ling et~al\mbox{.}(2021)]%
        {ling2021secure}
\bibfield{author}{\bibinfo{person}{Zhen Ling}, \bibinfo{person}{Huaiyu Yan}, \bibinfo{person}{Xinhui Shao}, \bibinfo{person}{Junzhou Luo}, \bibinfo{person}{Yiling Xu}, \bibinfo{person}{Bryan Pearson}, {and} \bibinfo{person}{Xinwen Fu}.} \bibinfo{year}{2021}\natexlab{}.
\newblock \showarticletitle{Secure boot, trusted boot and remote attestation for ARM TrustZone-based IoT Nodes}.
\newblock \bibinfo{journal}{\emph{Journal of Systems Architecture}} (\bibinfo{year}{2021}).
\newblock


\bibitem[Mamdouh et~al\mbox{.}(2018)]%
        {mamdouh2018securing}
\bibfield{author}{\bibinfo{person}{Marwa Mamdouh}, \bibinfo{person}{Mohamed~AI Elrukhsi}, {and} \bibinfo{person}{Ahmed Khattab}.} \bibinfo{year}{2018}\natexlab{}.
\newblock \showarticletitle{Securing the internet of things and wireless sensor networks via machine learning: A survey}. In \bibinfo{booktitle}{\emph{2018 International Conference on Computer and Applications (ICCA)}}.
\newblock


\bibitem[McCune et~al\mbox{.}(2010)]%
        {mccune2010trustvisor}
\bibfield{author}{\bibinfo{person}{Jonathan~M McCune}, \bibinfo{person}{Yanlin Li}, \bibinfo{person}{Ning Qu}, \bibinfo{person}{Zongwei Zhou}, \bibinfo{person}{Anupam Datta}, \bibinfo{person}{Virgil Gligor}, {and} \bibinfo{person}{Adrian Perrig}.} \bibinfo{year}{2010}\natexlab{}.
\newblock \showarticletitle{TrustVisor: Efficient TCB reduction and attestation}. In \bibinfo{booktitle}{\emph{2010 IEEE Symposium on Security and Privacy}}.
\newblock


\bibitem[McMillan and McMillan(1993)]%
        {mcmillan1993smv}
\bibfield{author}{\bibinfo{person}{Kenneth~L McMillan} {and} \bibinfo{person}{Kenneth~L McMillan}.} \bibinfo{year}{1993}\natexlab{}.
\newblock \showarticletitle{The SMV system}.
\newblock \bibinfo{journal}{\emph{Symbolic Model Checking}} (\bibinfo{year}{1993}), \bibinfo{pages}{61--85}.
\newblock


\bibitem[Muraleedharan and Osadciw(2006)]%
        {muraleedharan2006jamming}
\bibfield{author}{\bibinfo{person}{Rajani Muraleedharan} {and} \bibinfo{person}{Lisa~Ann Osadciw}.} \bibinfo{year}{2006}\natexlab{}.
\newblock \showarticletitle{Jamming attack detection and countermeasures in wireless sensor network using ant system}. In \bibinfo{booktitle}{\emph{Wireless Sensing and Processing}}.
\newblock


\bibitem[Noorman et~al\mbox{.}(2013)]%
        {noorman2013sancus}
\bibfield{author}{\bibinfo{person}{Job Noorman}, \bibinfo{person}{Pieter Agten}, \bibinfo{person}{Wilfried Daniels}, \bibinfo{person}{Raoul Strackx}, \bibinfo{person}{Anthony Van~Herrewege}, \bibinfo{person}{Christophe Huygens}, \bibinfo{person}{Bart Preneel}, \bibinfo{person}{Ingrid Verbauwhede}, {and} \bibinfo{person}{Frank Piessens}.} \bibinfo{year}{2013}\natexlab{}.
\newblock \showarticletitle{Sancus: Low-cost trustworthy extensible networked devices with a zero-software trusted computing base}. In \bibinfo{booktitle}{\emph{22nd USENIX Security Symposium (USENIX Security 13)}}.
\newblock


\bibitem[Nunes et~al\mbox{.}(2019a)]%
        {nunes2019towards}
\bibfield{author}{\bibinfo{person}{Ivan De~Oliveira Nunes}, \bibinfo{person}{Ghada Dessouky}, \bibinfo{person}{Ahmad Ibrahim}, \bibinfo{person}{Norrathep Rattanavipanon}, \bibinfo{person}{Ahmad-Reza Sadeghi}, {and} \bibinfo{person}{Gene Tsudik}.} \bibinfo{year}{2019}\natexlab{a}.
\newblock \showarticletitle{Towards systematic design of collective remote attestation protocols}. In \bibinfo{booktitle}{\emph{2019 IEEE 39th International Conference on Distributed Computing Systems (ICDCS)}}.
\newblock


\bibitem[Nunes et~al\mbox{.}(2019b)]%
        {vrased}
\bibfield{author}{\bibinfo{person}{Ivan De~Oliveira Nunes}, \bibinfo{person}{Karim Eldefrawy}, \bibinfo{person}{Norrathep Rattanavipanon}, \bibinfo{person}{Michael Steiner}, {and} \bibinfo{person}{Gene Tsudik}.} \bibinfo{year}{2019}\natexlab{b}.
\newblock \showarticletitle{$\{$VRASED$\}$: A verified $\{$Hardware/Software$\}$$\{$Co-Design$\}$ for remote attestation}. In \bibinfo{booktitle}{\emph{28th USENIX Security Symposium (USENIX Security 19)}}.
\newblock


\bibitem[Nunes et~al\mbox{.}(2020)]%
        {nunes2020apex}
\bibfield{author}{\bibinfo{person}{Ivan De~Oliveira Nunes}, \bibinfo{person}{Karim Eldefrawy}, \bibinfo{person}{Norrathep Rattanavipanon}, {and} \bibinfo{person}{Gene Tsudik}.} \bibinfo{year}{2020}\natexlab{}.
\newblock \showarticletitle{$\{$APEX$\}$: A verified architecture for proofs of execution on remote devices under full software compromise}. In \bibinfo{booktitle}{\emph{29th USENIX Security Symposium (USENIX Security 20)}}.
\newblock


\bibitem[Obermaier and Immler(2018)]%
        {obermaier2018past}
\bibfield{author}{\bibinfo{person}{Johannes Obermaier} {and} \bibinfo{person}{Vincent Immler}.} \bibinfo{year}{2018}\natexlab{}.
\newblock \showarticletitle{The past, present, and future of physical security enclosures: from battery-backed monitoring to puf-based inherent security and beyond}.
\newblock \bibinfo{journal}{\emph{Journal of hardware and systems security}} (\bibinfo{year}{2018}).
\newblock


\bibitem[{Olivier Girard}(2009)]%
        {openMSP430}
\bibfield{author}{\bibinfo{person}{{Olivier Girard}}.} \bibinfo{year}{2009}\natexlab{}.
\newblock \bibinfo{title}{OpenMSP430}.
\newblock \bibinfo{howpublished}{\url{https://opencores.org/projects/openmsp430/}}.
\newblock


\bibitem[{OpenSim Ltd}({[n.\,d.]})]%
        {omnetpp}
\bibfield{author}{\bibinfo{person}{{OpenSim Ltd}}.} \bibinfo{year}{[n.\,d.]}\natexlab{}.
\newblock \bibinfo{title}{OMNeT++ Discrete Event Simulator}.
\newblock \bibinfo{howpublished}{\url{https://omnetpp.org/}}.
\newblock


\bibitem[{P. Frolikov, Y. Kim, R. Prapty, G. Tsudik}({[n.\,d.]})]%
        {TRAPSAnonOpenSource}
\bibfield{author}{\bibinfo{person}{{P. Frolikov, Y. Kim, R. Prapty, G. Tsudik}}.} \bibinfo{year}{[n.\,d.]}\natexlab{}.
\newblock \bibinfo{title}{TRAIN source code}.
\newblock \bibinfo{howpublished}{\url{https://github.com/sprout-uci/TRAIN}}.
\newblock


\bibitem[Perrig et~al\mbox{.}(2003)]%
        {perrig-tesla}
\bibfield{author}{\bibinfo{person}{Adrian Perrig}, \bibinfo{person}{JD Tygar}, \bibinfo{person}{Adrian Perrig}, {and} \bibinfo{person}{JD Tygar}.} \bibinfo{year}{2003}\natexlab{}.
\newblock \showarticletitle{TESLA broadcast authentication}.
\newblock \bibinfo{journal}{\emph{Secure Broadcast Communication: In Wired and Wireless Networks}} (\bibinfo{year}{2003}).
\newblock


\bibitem[Petzi et~al\mbox{.}(2022)]%
        {petzi2022scraps}
\bibfield{author}{\bibinfo{person}{Lukas Petzi}, \bibinfo{person}{Ala Eddine~Ben Yahya}, \bibinfo{person}{Alexandra Dmitrienko}, \bibinfo{person}{Gene Tsudik}, \bibinfo{person}{Thomas Prantl}, {and} \bibinfo{person}{Samuel Kounev}.} \bibinfo{year}{2022}\natexlab{}.
\newblock \showarticletitle{$\{$SCRAPS$\}$: Scalable Collective Remote Attestation for $\{$Pub-Sub$\}$$\{$IoT$\}$ Networks with Untrusted Proxy Verifier}. In \bibinfo{booktitle}{\emph{31st USENIX Security Symposium (USENIX Security 22)}}.
\newblock


\bibitem[Ravi et~al\mbox{.}(2004)]%
        {ravi2004tamper}
\bibfield{author}{\bibinfo{person}{Srivaths Ravi}, \bibinfo{person}{Anand Raghunathan}, {and} \bibinfo{person}{Srimat Chakradhar}.} \bibinfo{year}{2004}\natexlab{}.
\newblock \showarticletitle{Tamper resistance mechanisms for secure embedded systems}. In \bibinfo{booktitle}{\emph{VLSI Design}}.
\newblock


\bibitem[Sehatbakhsh et~al\mbox{.}(2019)]%
        {sehatbakhsh2019emma}
\bibfield{author}{\bibinfo{person}{Nader Sehatbakhsh}, \bibinfo{person}{Alireza Nazari}, \bibinfo{person}{Haider Khan}, \bibinfo{person}{Alenka Zajic}, {and} \bibinfo{person}{Milos Prvulovic}.} \bibinfo{year}{2019}\natexlab{}.
\newblock \showarticletitle{Emma: Hardware/software attestation framework for embedded systems using electromagnetic signals}. In \bibinfo{booktitle}{\emph{Proceedings of the 52nd Annual IEEE/ACM International Symposium on Microarchitecture}}.
\newblock


\bibitem[Seshadri et~al\mbox{.}(2006)]%
        {seshadri2006scuba}
\bibfield{author}{\bibinfo{person}{Arvind Seshadri}, \bibinfo{person}{Mark Luk}, \bibinfo{person}{Adrian Perrig}, \bibinfo{person}{Leendert Van~Doorn}, {and} \bibinfo{person}{Pradeep Khosla}.} \bibinfo{year}{2006}\natexlab{}.
\newblock \showarticletitle{SCUBA: Secure code update by attestation in sensor networks}. In \bibinfo{booktitle}{\emph{Proceedings of the 5th ACM workshop on Wireless security}}.
\newblock


\bibitem[Seshadri et~al\mbox{.}(2004)]%
        {seshadri2004swatt}
\bibfield{author}{\bibinfo{person}{Arvind Seshadri}, \bibinfo{person}{Adrian Perrig}, \bibinfo{person}{Leendert Van~Doorn}, {and} \bibinfo{person}{Pradeep Khosla}.} \bibinfo{year}{2004}\natexlab{}.
\newblock \showarticletitle{SWATT: Software-based attestation for embedded devices}. In \bibinfo{booktitle}{\emph{IEEE Symposium on Security and Privacy, 2004. Proceedings. 2004}}.
\newblock


\bibitem[Strackx et~al\mbox{.}(2010)]%
        {strackx2010efficient}
\bibfield{author}{\bibinfo{person}{Raoul Strackx}, \bibinfo{person}{Frank Piessens}, {and} \bibinfo{person}{Bart Preneel}.} \bibinfo{year}{2010}\natexlab{}.
\newblock \showarticletitle{Efficient isolation of trusted subsystems in embedded systems}. In \bibinfo{booktitle}{\emph{Security and Privacy in Communication Networks: 6th Iternational ICST Conference, SecureComm 2010, Singapore, September 7-9, 2010. Proceedings 6}}.
\newblock


\bibitem[Surminski et~al\mbox{.}(2021)]%
        {surminski2021realswatt}
\bibfield{author}{\bibinfo{person}{Sebastian Surminski}, \bibinfo{person}{Christian Niesler}, \bibinfo{person}{Ferdinand Brasser}, \bibinfo{person}{Lucas Davi}, {and} \bibinfo{person}{Ahmad-Reza Sadeghi}.} \bibinfo{year}{2021}\natexlab{}.
\newblock \showarticletitle{Realswatt: Remote software-based attestation for embedded devices under realtime constraints}. In \bibinfo{booktitle}{\emph{Proceedings of the 2021 ACM SIGSAC Conference on Computer and Communications Security}}.
\newblock


\bibitem[Technology({[n.\,d.]})]%
        {rtc}
\bibfield{author}{\bibinfo{person}{Microchip Technology}.} \bibinfo{year}{[n.\,d.]}\natexlab{}.
\newblock \bibinfo{title}{MCP7940M: Low-Cost I2C Real-Time Clock/Calendar with SRAM}.
\newblock
\newblock
\urldef\tempurl%
\url{https://ww1.microchip.com/downloads/en/DeviceDoc/MCP7940M-Low-Cost%%20I2C-RTCC-with-SRAM-20002292C.pdf}
\showURL{%
\tempurl}


\bibitem[{Texas Instruments}(2016)]%
        {msp430-gcc}
\bibfield{author}{\bibinfo{person}{{Texas Instruments}}.} \bibinfo{year}{2016}\natexlab{}.
\newblock \bibinfo{title}{{MSP430} {GCC} User's Guide}.
\newblock \bibinfo{howpublished}{\url{https://www.ti.com/tool/MSP430-GCC-OPENSOURCE/}}.
\newblock


\bibitem[Wang et~al\mbox{.}(2023)]%
        {wang2023ari}
\bibfield{author}{\bibinfo{person}{Jinwen Wang}, \bibinfo{person}{Yujie Wang}, \bibinfo{person}{Ao Li}, \bibinfo{person}{Yang Xiao}, \bibinfo{person}{Ruide Zhang}, \bibinfo{person}{Wenjing Lou}, \bibinfo{person}{Y~Thomas Hou}, {and} \bibinfo{person}{Ning Zhang}.} \bibinfo{year}{2023}\natexlab{}.
\newblock \showarticletitle{$\{$ARI$\}$: Attestation of Real-time Mission Execution Integrity}. In \bibinfo{booktitle}{\emph{32nd USENIX Security Symposium (USENIX Security 23)}}.
\newblock


\bibitem[Zhijun et~al\mbox{.}(2020)]%
        {zhijun2020low}
\bibfield{author}{\bibinfo{person}{Wu Zhijun}, \bibinfo{person}{Li Wenjing}, \bibinfo{person}{Liu Liang}, {and} \bibinfo{person}{Yue Meng}.} \bibinfo{year}{2020}\natexlab{}.
\newblock \showarticletitle{Low-rate DoS attacks, detection, defense, and challenges: A survey}.
\newblock \bibinfo{journal}{\emph{IEEE access}} (\bibinfo{year}{2020}).
\newblock


\end{thebibliography}

\end{document}